%% file: paper.tex
\documentclass[a4paper,oneside,11pt]{article}
\usepackage{cprform}
\usepackage{amsmath,amstext,amsfonts,amsbsy,amssymb,amscd,bbm,epsfig,lscape}

\usepackage{epsfig,cite}
\usepackage{amssymb}
\usepackage{subfigure}

\input{macros.tex}

\begin{document}

\begin{titlepage}


\vspace*{-30truemm}
\begin{flushright}
CERN-PH-TH/2007-252\\
ROM2F/2007/20\\
DESY 07-220\\
FTUAM-07-20\\
IFT-UAM-CSIC-07-64\\
MKPH-T-07-25\\
\vspace{5truemm}
{\large December 2007}
\end{flushright}
\vspace{5truemm}

\vskip -0.0cm
\begin{center}
{\Bigrm Non-perturbative renormalisation of {\Large ${\Delta F=2}$} }\\
{\Bigrm four-fermion operators in two-flavour QCD}
\end{center}
\vskip 4 true mm
\begin{center}
\epsfig{figure=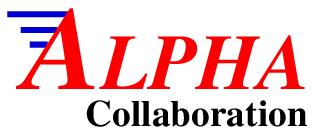, width=22 true mm}\\
\end{center}
\vskip -2 true mm
\centerline{\bigrm
P.~Dimopoulos$^a$, 
G.~Herdoiza$^b$,
F.~Palombi$^c$,}
\centerline{\bigrm
M.~Papinutto$^c$,
C.~Pena$^d$, 
A.~Vladikas$^a$
and H.~Wittig$^e$}
\vskip 4 true mm
\centerline{\it $^a$ INFN, Sezione di Roma II}
\centerline{\it and Dipartimento di Fisica, Universit\`a di Roma ``Tor Vergata''}
\centerline{\it Via della Ricerca Scientifica 1, I-00133 Rome, Italy}
\vskip 3 true mm
\centerline{\it $^b$ DESY, Platanenallee 6, D-15738 Zeuthen, Germany}
\vskip 3 true mm
\centerline{\it $^c$ CERN, Physics Department, TH Division, CH-1211 Geneva, Switzerland}
\vskip 3 true mm
\centerline{\it $^d$ Departamento de F\'{\i}sica Te\'orica C-XI and}
\centerline{\it Instituto de F\'{\i}sica Te\'orica UAM/CSIC C-XVI}
\centerline{\it Universidad Aut\'onoma de Madrid, Cantoblanco E-28049 Madrid, Spain}
\vskip 3 true mm
\centerline{\it $^e$ Johannes Gutenberg Universit\"at, Institut f\"ur Kernphysik}
\centerline{\it Johann Joachim Becher-Weg 45, D-55099 Mainz, Germany}
\vskip 3 true mm
\vskip 5 true mm


\thicktablerule
\vskip 3 true mm
\noindent{\tenbf Abstract}
\vskip 1 true mm
\noindent
Using Schr\"odinger Functional methods, we compute the non-perturbative 
renormalisation and renormalisation group running of several four-fermion 
operators, in the framework of lattice simulations with two dynamical 
Wilson quarks. Two classes of operators have been
targeted: (i)  those with left-left current structure and four propagating 
quark fields; (ii) all operators containing two static quarks. In both 
cases, only the parity-odd contributions have been considered, being the 
ones that renormalise multiplicatively. Our results, once combined with 
future simulations of the corresponding lattice hadronic matrix elements, 
may be used for the computation of phenomenological quantities of interest, 
such as $B_K$ and $B_B$ (the latter also in the static limit).
{\tenrm }
\vskip 3 true mm
\thicktablerule
\vspace{10truemm}
\eject

\end{titlepage}

\section{Introduction}

Hadronic matrix elements of four-fermion operators play an important r\^ole
in the study of CP violation via CKM  unitarity triangle analyses, as well as in the
understanding of the $\Delta I=1/2$ enhancement puzzle in $K \rightarrow \pi\pi$ decays.
The only known technique to compute hadronic matrix elements from first principles, 
namely lattice QCD, has long been hampered by a number of systematic uncertainties. 
Most notably, the high computational cost of including light dynamical quarks in the 
simulations has enforced either the quenched  approximation, or the use of heavy 
dynamical quark masses, which necessitate long and potentially uncontrolled extrapolations
to the chiral regime. It is thus important to upgrade existing quenched results by the 
inclusion of dynamical fermion effects. For recent progress reports on lattice results
on flavour Physics, see~\cite{Juttner:2007sn}.

The present work is a step in this direction. The non-perturbative renormalisation of 
four-fermion operators, as well as the corresponding renormalisation group (RG) running 
between hadronic scales of ${\rm O}(\Lambda_{\rm QCD})$ and perturbative ones of about 
$100$~GeV, is a necessary ingredient in the process of producing the properly 
renormalised matrix element in the continuum limit. Quantities that determine these 
renormalisation properties have been computed in the quenched approximation, using 
finite-size scaling techniques, for a broad class of  four-fermion 
operators~\cite{Guagnelli:2005zc,Palombi:2005zd,Palombi:2007dr}. 
The regularisation of choice was that of non-perturbatively ${\rm O}(a)$ improved Wilson quarks and
standard plaquette gauge action; the renormalisation schemes used were of the Schr\"odinger functional (SF) type.
The aim of the present work is to extend these results to QCD with $\NF=2$
dynamical quarks. More specifically, we present results for: (i) the RG-running of
left-left current relativistic four-fermion operators; (ii) the RG-running of all
$\Delta B=2$ operators with two static heavy quarks; (iii) the renormalisation factors that
match the above operators to their renormalisation group invariant (RGI) counterparts. 
The latter have been computed for a regularisation of the relativistic quarks by the 
non-perturbatively ${\rm O}(a)$ improved Wilson action. Preliminary results
have been presented in~\cite{Dimopoulos:2006es}.

As we will point out below, the results of this work are relevant for the computation
of physical quantities such as the bag parameters $B_K$ and $B_B$. 
In the quenched approximation, the combination of this renormalisation programme
with computations  of bare hadronic matrix elements has already produced high precision estimates of
a few physical quantities in the continuum~\cite{Dimopoulos:2006dm,Dimopoulos:2007cn}.
Moreover, knowledge of the continuum RGI operators, computed with Wilson fermions,
has allowed the determination, through a matching procedure, of the renormalisation 
factors of the same operators in the Neuberger fermion regularisation~\cite{Dimopoulos:2006ma}.

The paper is organised as follows. In sect.~2 we introduce the operator basis and the
renormalisation schemes adopted in the present work. We also recall some basic
formulae used for the reconstruction of the operator scale evolution in the SF framework. 
Sect.~3 is devoted to a detailed description of the lattice simulations and numerical 
analyses of the operator RG-running. In sect.~4 we discuss the renormalisation of the
four-quark operators at a low energy matching scale. Conclusions are drawn in sect.~5. 
In order to improve readability, some tables and figures have been collected at the 
end of the paper. 

\section{Definitions and setup}

\subsection{Renormalisation of four-fermion operators}

We will consider two different classes of four-fermion (dimension-six) operators:
\begin{align}
O^\pm_{\Gamma_1\Gamma_2}(x) &= \frac{1}{2}\left[
 \left(\bar\psi_1(x)\Gamma_1\psi_2(x)\right)
 \left(\bar\psi_3(x)\Gamma_2\psi_4(x)\right)
 \pm ( \psi_2\leftrightarrow \psi_4) \right] \ , \\
\cO^\pm_{\Gamma_1\Gamma_2}(x) &= \frac{1}{2}\left[
 \left(\heavyb(x)\Gamma_1\psi_1(x)\right)
 \left(\aheavyb(x)\Gamma_2\psi_2(x)\right)
 \pm ( \psi_1\leftrightarrow \psi_2) \right] \ .
\end{align}
In the above expressions $\psi_k$ is a relativistic quark field with
flavour index $k$, $\heavy$ ($\aheavy$) are static (anti-)quark fields,
$\Gamma_l$ are Dirac (spin) matrices, and the parentheses indicate 
summation over spin and colour indices. In the present formalism, all 
quark flavours are distinct, enabling us to separate the calculation 
of the scale-dependent logarithmic divergences, which is the aim of 
the present work, from the problem of eventual mixing with 
lower-dimensional operators\footnote{These power subtractions typically 
appear for some specific choices of quark masses and/or flavour content 
(e.g. penguin operators). Their determination is independent of
that of the logarithmic divergences, once mass independent renormalisation
schemes are employed.}.

The renormalisation pattern of the above operators is determined by the
symmetries of the regularised theory. In the the parity-odd sector,
complete bases of operators in the relativistic and static cases are
given by
\begin{alignat}{3}
Q^\pm_k & \in \left\{
O^\pm_{\rm\scriptscriptstyle VA+AV},
O^\pm_{\rm\scriptscriptstyle VA-AV},
O^\pm_{\rm\scriptscriptstyle SP-PS},
O^\pm_{\rm\scriptscriptstyle SP+PS},
O^\pm_{\rm\scriptscriptstyle T \tilde T},
\right\} \ , & & & \qquad k = 1,\dots,5 \ , \\[2.0ex]
\cQ^\pm_k & \in \left\{
\cO^\pm_{\rm\scriptscriptstyle VA+AV},
\cO^\pm_{\rm\scriptscriptstyle VA-AV},
\cO^\pm_{\rm\scriptscriptstyle SP-PS},
\cO^\pm_{\rm\scriptscriptstyle SP+PS},
\right\} \ , \qquad & & & k = 1,\dots,4\ , 
\end{alignat}
respectively. The notation is standard and self-explanatory, indicating the operator spin 
matrices $\Gamma_l$, with say, $O^\pm_{\rm\scriptscriptstyle VA+AV} \equiv
O^\pm_{\rm\scriptscriptstyle VA} + O^\pm_{\rm\scriptscriptstyle AV}$. 
A full analysis of the renormalisation properties of these
operator bases with relativistic Wilson fermions has been performed in~\cite{Donini:1999sf,Palombi:2006pu}. 
A result of these works which is of particular relevance is that, contrary to the parity-even case, 
characterised by operator mixing due to the explicit breaking of chiral symmetry by the Wilson term in
the quark action, the parity-odd operators are protected by discrete symmetries, 
and hence their renormalisation pattern is continuum-like~\cite{Bernard:1987pr}. 
We point out that RG-running is identical for parity-even and parity-odd operators
of the same chiral representation, since in the continuum limit chiral symmetry is restored.
On the other hand, the (physically relevant) matrix elements of
the parity-even operators can be mapped exactly to those of the
parity-odd ones via the addition of a chirally twisted mass
term to the lattice quark action~\cite{Frezzotti:2000nk,Dimopoulos:2006dm,Palombi:2006pu}.

From now on, we will consider the subset of operators
\begin{gather}
\label{operbasis}
Q_1^\pm \ ,~~
{\cQ'}^{+}_k \in \left\{ 
\cQ^+_1,\cQ^+_1+4\cQ^+_2,\cQ^+_3+2\cQ^+_4,\cQ^+_3-2\cQ^+_4
\right\} \ .
\end{gather}
All these operators renormalise multiplicatively; i.e. given
an operator $O\in\{Q_1^\pm,{\cQ'}_k^{+}\}$ the corresponding operator
insertion in any on-shell renormalised correlation function is given by
\begin{gather}
O_{\rm\scriptscriptstyle R}(x,\mu) = \lim_{a \to 0} Z(g_0,a\mu)\,O(x;g_0) \ ,
\end{gather}
where $g_0,a$ are the bare coupling and lattice spacing,
respectively and $\mu$ is the
renormalisation scale. The RG-running of the operator is controlled by the
anomalous dimension $\gamma$, defined by the Callan-Symanzik equation
\begin{gather}
\label{eq:CS}
\mu\frac{\partial}{\partial\mu}O_{\rm\scriptscriptstyle R}(x,\mu) = \gamma\left(\gbar(\mu)\right)\,O_{\rm\scriptscriptstyle R}(x,\mu) \ ,
\end{gather}
supplemented by the corresponding RG-equation
for the renormalised coupling $\gbar$,
\begin{gather}
\mu\frac{\partial}{\partial\mu}\gbar(\mu) = \beta(\gbar(\mu))\ .
\end{gather}
In mass-independent renormalisation schemes, the $\beta$-function and all
anomalous dimensions depend only on the renormalised coupling 
$\gbar$. They admit perturbative expansions of the form
\begin{align}
\beta(g) & \stackrel{g\to 0}{\approx} -g^3\left(b_0+b_1g^2+b_2g^4+\ldots\right) \ , \label{betafunction}\\[1.5ex]
\gamma(g) & \stackrel{g\to 0}{\approx} -g^2\left(
\gamma_0 + \gamma_1g^2 + \gamma_2g^4 + \ldots\right) \ , \label{anomalousdimen}
\end{align}
in which the coefficients $b_0,b_1,\gamma_0$ are renormalisation scheme-independent. 
In particular, the universal coefficients of the $\beta$-function read
\begin{equation}
b_0 = \frac{1}{(4\pi)^2}\biggl\{11 - \frac{2}{3}N_{\rm f}\biggr\}    \ , 
\qquad b_1 = \frac{1}{(4\pi)^4}\biggl\{102 - \frac{38}{3}N_{\rm f}\biggr\}\ \ ,
\end{equation}
and the universal leading order (LO) coefficients of the anomalous dimensions of the 
operators in Eq.~(\ref{operbasis}) are given by
\begin{alignat}{4}
\gamma_0^+ & = \phantom{-}\frac{4}{(4\pi)^2}\ , \qquad &\gamma_0^- & = -\frac{8}{(4\pi)^2}\ ,  \\[0.6ex]
\gamma_0^{(1)} & = -\frac{8}{(4\pi)^2}\ , \qquad & \gamma_0^{(2)} & = -\frac{8}{3(4\pi)^2}\ ,      \\[0.6ex]
\gamma_0^{(3)} & = -\frac{10}{(4\pi)^2}\ , \qquad & \gamma_0^{(4)} & = -\frac{4}{(4\pi)^2}\ .       
\end{alignat}
Moreover, in the SF renormalisation scheme, the next-to-next-to-leading order (NNLO) coefficient 
$b_2^{\rm\scriptscriptstyle SF}$ of the $\beta$-function is known to be~\cite{DellaMorte:2004bc}
\begin{equation}
b_2^{\rm\scriptscriptstyle SF} = \frac{1}{(4\pi)^3}\biggl\{0.483 - 0.275N_{\rm f} + 0.0361N_{\rm f}^2 - 0.00175N_{\rm f}^3\biggr\}\ .
\end{equation}

Upon formal integration of Eq.~(\ref{eq:CS}), one obtains the renormalisation 
group invariant (RGI) operator insertion
\begin{gather}
\label{eq:rgi}
O_{\rm\scriptscriptstyle RGI}(x) = O_{\rm\scriptscriptstyle R}(x;\mu)\,
\left[\frac{\gbar^2(\mu)}{4\pi}\right]^{-\frac{\gamma_0}{2b_0}}
\exp\left\{
-\int_0^{\gbar(\mu)}\dif g\,\left(\frac{\gamma(g)}{\beta(g)}-
\frac{\gamma_0}{b_0g}\right)
\right\} \ ,
\end{gather}
while the RG evolution between two scales $\mu_1,\mu_2$ is given
by the scaling factor
\begin{gather}
\label{eq:evol}
U(\mu_2,\mu_1) = \exp\left\{
\int_{\gbar(\mu_1)}^{\gbar(\mu_2)}
\dif g \, \frac{\gamma(g)}{\beta(g)}
\right\} = \lim_{a \to 0} \,\frac{Z(g_0,a\mu_2)}{Z(g_0,a\mu_1)} \ .
\end{gather}

\subsection{Schr\"odinger Functional renormalisation schemes}

Eqs.~(\ref{eq:rgi})-(\ref{eq:evol}) are the starting point for the 
non-perturbative computation of the RG evolution of composite operators. 
To that purpose we introduce a family of Schr\"odinger Functional 
renormalisation schemes. The latter are defined by setting up the theory 
on a four-dimensional hypercube of physical size $T\times L^3$ with 
Dirichlet boundary conditions in Euclidean time and periodic boundary 
conditions in the spatial directions, up to a phase $\theta$. We refer 
the reader to \cite{Luscher:1996sc} for an introduction to the SF setup. 
In the present work we always choose $T=L$ and $\theta=0.5$. We also 
assume that no background field is present. The renormalisation scale 
is set as $\mu=1/L$.

Renormalisation conditions are imposed on SF correlators,
following~\cite{Guagnelli:2005zc,Palombi:2005zd,Palombi:2007dr}; for the sake
of completeness, we briefly outline the method.
We first introduce bilinear boundary sources projected to zero external momentum,
\begin{align}
\Sigma_{s_1s_2}[\Gamma]  & = a^6 \sum_{\bx\by}\bar\zeta_{s_1}(\bx)\Gamma\zeta_{s_2}(\by)\ , \\[1.5ex]
\Sigma'_{s_1s_2}[\Gamma] & = a^6 \sum_{\bx\by}\bar\zeta'_{s_1}(\bx)\Gamma\zeta'_{s_2}(\by)\ .
\end{align}
Here $\Gamma$ denotes a Dirac matrix, the flavour indices $s_{1,2}$ can assume both
relativistic and static values and the fields $\zeta$ ($\zeta'$) represent functional derivatives
with respect to the fermionic boundary fields of the SF at the initial (final) time $x_0 = 0$ ($x_0 = T$). 
The four-quark operators are then treated as local insertions in the bulk of the SF, giving rise to
the correlation functions
\begin{align}
F^\pm_{[\Gamma_{\rm A},\Gamma_{\rm B},\Gamma_{\rm C}]}(x_0) & = L^{-3}\langle \Sigma'_{53}[\Gamma_{\rm C}] \,\,\, Q_1^\pm(x) \,\, \Sigma_{21}[\Gamma_{\rm A}] \,\, \Sigma_{45}[\Gamma_{\rm B}]  
\rangle\ , \label{relcorr}\\[1.5ex]
{\cal F}_{[\Gamma_{\rm A},\Gamma_{\rm B},\Gamma_{\rm C}]}^{(k)}(x_0) & = L^{-3}\langle \Sigma'_{3\rm \phantom{\bar{h}^\dagger}\hspace{-0.28cm}\bar{h}}[\Gamma_{\rm C}] \,\, {\cal Q'}_k^+(x) \,\, \Sigma_{1{\rm h}}[\Gamma_{\rm A}] \,\, \Sigma_{23}[\Gamma_{\rm B}] 
\rangle\ . \label{statcorr}
\end{align}
Clearly, the Dirac matrices of the boundary sources must be chosen so that the correlators
do not vanish trivially (e.g. due to parity).

In the above definitions a ``spectator'' light quark has been introduced with flavour
$s=5$ for $F^\pm$ and $s=3$ for ${\cal F}^{(k)}$. This quark field has no Wick contractions
with the valence quarks of the operator insertion and propagates straight from the initial to the
final time boundary. Its r\^ole is merely to allow for parity-even correlators of parity odd 
four-fermion insertions without the need of introducing non-zero external momenta.
\begin{figure}[!t]
\begin{center}
\epsfig{figure=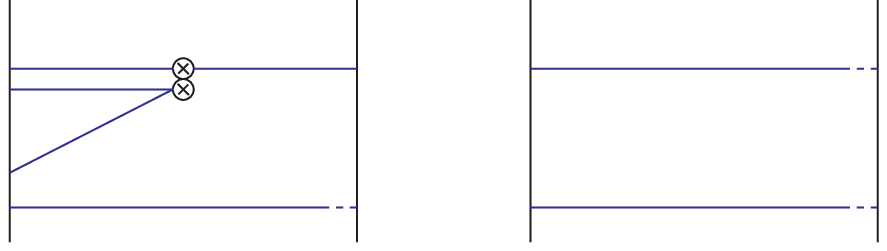, width=5.5 true cm} \hskip 1.0cm
\epsfig{figure=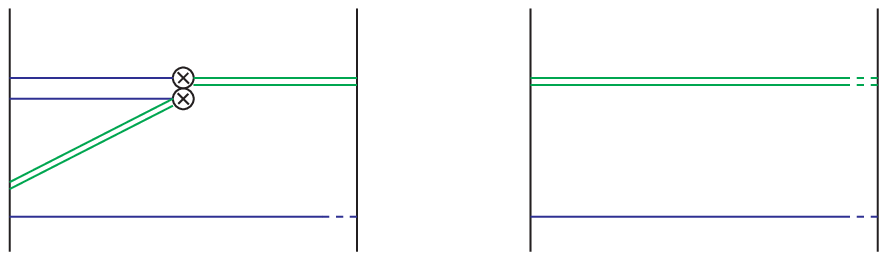, width=5.5 true cm}
\end{center}
\caption{\small Diagrammatic representation of correlation functions: relativistic four-quark correlators
$F^{\pm}$ (first diagram from left), static four-quark correlators ${\cal F}^{(k)}$ (third diagram from left),  
relativistic boundary-to-boundary correlators $f_1^{12}$, $k_1^{12}$ (second diagram from left) and
static boundary-to-boundary correlators $f_1^{1\rm h}$ and $k_1^{1\rm h}$ (fourth diagram from left). 
Euclidean time goes from left to right. Single (double) lines represent relativistic (static) valence
quarks.}
\label{fig:feynman}
\end{figure}

In order to isolate the operator ultraviolet divergences in Eqs.~(\ref{relcorr})-(\ref{statcorr}),
one has to remove the boundary sources' additional divergences. To this end, we introduce
a set of boundary-to-boundary correlators, 
\begin{align}
f_1^{s_1s_2} & = -\frac{1}{2L^6}\langle \Sigma'_{s_1s_2}[\gamma_5] \,\, \Sigma_{s_2s_1}[\gamma_5]\rangle\ , \\[2.0ex]
k_1^{s_1s_2} & = -\frac{1}{6L^6}\sum_{k=1}^3\langle \Sigma'_{s_1s_2}[\gamma_k] \,\, \Sigma_{s_2s_1}[\gamma_k]\rangle \ , 
\end{align}
where the flavour indices $s_{1,2}$ may assume once again either relativistic or static values. 
Wick contractions of four-quark and boundary-to-boundary correlation functions are depicted in 
Fig.~\ref{fig:feynman}. 

Since the logarithmic divergences of the boundary fields are removed by multiplicative 
renormalisation factors $Z_\zeta$ and $Z_{\zeta}^{\rm h}$, it can be easily recognised 
that ratios of correlators such as
\begin{align}
R^\pm(x_0) & = \frac{F^\pm_{[\Gamma_{\rm A},\Gamma_{\rm B},\Gamma_{\rm C}]}(x_0) }{[f_1^{12}]^{3/2-\alpha}[k_1^{12}]^\alpha}\ , \\[1.5ex]
{\cal R}^{(k)}(x_0) & = \frac{{\cal F}^{(k)}_{[\Gamma_{\rm A},\Gamma_{\rm B},\Gamma_{\rm C}]}(x_0) }{[f_1^{1{\rm h}}][f_1^{12}]^{1/2-\alpha}[k_1^{12}]^\alpha}\ ,
\end{align}
are free of boundary divergences for any choice of the parameter $\alpha$. Thus, the operators
of interest are renormalised through the conditions
\begin{align}
Z^\pm(g_0,a/L)R^\pm(T/2) & = R^\pm(T/2) \Big \vert_{g_0=0}\ , \label{relrenscheme}\\[2.0ex]
{\cal Z}^{(k)}(g_0,a/L){\cal R}^{(k)}(T/2) & = {\cal R}^{(k)}(T/2) \Big \vert_{g_0=0}\ , \label{statrenscheme}
\end{align}
where all correlation functions are to be evaluated at the chiral point.

A crucial observation is that all renormalisation factors thus obtained
are flavour-blind, in the sense that they remove the logarithmic divergences 
from any four-fermion operator of a given Dirac structure, irrespective of
its specific flavour content.
For instance, in $\Delta H=2$ transitions, such as $B_{d(s)}-\bar B_{d(s)}$ 
oscillations in the static limit, one identifies $\psi_1=\psi_2=d$($s$)
as a down (strange) quark, and $\psi_3 = u$ as an up quark; for either flavour identification
the same ${\cal Z}^{(k)}$'s renormalise the corresponding operator. Similarly,
in the relativistic quark case, the dimension-six operator, be it $\Delta S=2$
(with $\psi_1 = \psi_3 = s$ and $\psi_2 = \psi_4 = d$) or $\Delta B=2$
(with $\psi_1 = \psi_3 = b$ and $\psi_2 = \psi_4 = d(s)$), is renormalised by 
the same $Z^+$. Also in this case we note the presence of the spectator quark
$\psi_5 = u$ in the renormalisation condition. These  renormalisation factors
also remove the logarithmic divergences of other dimension-six operators with the
same Dirac structure but different flavour content. For example,
even if the renormalisation of some $\Delta F = 1$ operators is only complete 
after power subtractions are taken into account, their logarithmic divergences
are removed by the same $Z^\pm$ and ${\cal Z}^{(k)}$ factors.

Another issue, related to the above discussion, is that here we are
working with $\NF=2$ dynamical flavours. In large-volume simulations of 
hadronic matrix elements, the latter would be naturally identified with
the up  and down sea quarks. However, most matrix elements of interest
involve  additional propagating physical flavours\footnote{An exception
is  that of $\Delta B=2$ oscillations of the $B_d$ meson in the static
limit.}. Thus, it becomes necessary to address the effects of partial 
quenching, in the context of operator renormalisation.

Due to their flavour-blind nature, the renormalisation factors and RG
running  contained in this work account for the scale dependence of any 
matrix element of four-fermion operator, computed on an $\NF=2$ sea, 
provided that partial quenching does not  generate extra scale-dependent 
mixing, which spoils multiplicative renormalisation. This is indeed 
the case for $\Delta F=2$ meson oscillations, since in the chiral  limit 
the relevant symmetries (CPS for $Q_1^\pm$ and CPT plus heavy quark 
spin symmetry for ${\cQ'}_k^{+}$) remain valid and protect these operators 
from new counterterms. On the contrary, unphysical mixing, generated by 
partial quenching, may become an issue for $\Delta F=1$ hadronic decays, at 
least for penguin contributions \cite{Golterman:2001qj}. We stress that this 
problem is not specific to the  renormalisation schemes under consideration.

\subsection{Step-scaling functions}

The non-perturbative study of the RG-evolution of our composite operators
is based on the step-scaling functions (SSFs)
\begin{align}
\sigma(u) & = \lim_{a \to 0} \Sigma(u,a/L)\ , \qquad \Sigma(u,a/L) = \left.\frac{Z\left(g_0,a/(2L)\right)}{Z(g_0,a/L)}\right|_{\ \gbar^2_{\rm\scriptscriptstyle SF}(L)=u,\ m_{\scriptscriptstyle 0}=m_{\rm cr}}\ ,
\end{align}
with $Z \in \{Z^{\pm},{\cal Z}^{(k)}\}$. According to Eq.~(\ref{eq:evol}), the SSFs 
describe the operator RG-running between the scales $\mu_1 = 1/L$ and $\mu_2 = 1/(2L)$. 
The choice  of the ratio $\mu_1/\mu_2=2$ as the smallest positive integer is made with the aim
of minimising the effects of the ultraviolet cutoff at finite values of the latter. 

In practice, the SSFs are simulated at several values of the lattice spacing for fixed
physical size (inverse renormalisation scale) $L$. The corresponding values
of the inverse bare coupling $\beta=6/g_0^2$ are indeed tuned by requiring that 
the renormalised SF coupling $\gbar^2_{\rm\scriptscriptstyle SF}$, and hence $L$, are kept constant. 
The critical mass $m_{\rm cr}$ is obtained from the requirement that the PCAC mass vanishes
in the ${\rm O}(a)$ improved theory. 

Once $\sigma(u)$ is computed at several different values of the squared gauge coupling 
$u$, it is possible to reconstruct the RG evolution factor $U(\mu_{\rm pt},\mu_{\rm had})$ 
between a hadronic scale $\mu_{\rm had}$, in the range of a few hundred $\MeV$, and a 
perturbative one $\mu_{\rm pt}$, in the high-energy regime. This in turn
leads to the computation of the RGI operator of Eq.~(\ref{eq:rgi}), with controlled
systematic uncertainties, by splitting the exponential on the rhs of Eq.~(\ref{eq:rgi}),
evaluated at $\mu=\mu_{\rm had}$, as follows:
\begin{align}
\label{RGrunexp}
\hat c(\mu_{\rm had}) & \equiv \left[\frac{\gbar^2(\mu_{\rm had})}{4\pi}\right]^{-\frac{\gamma_0}{2b_0}}\exp\left\{-\int_0^{\gbar(\mu_{\rm had})}\dif g\,\left(\frac{\gamma(g)}{\beta(g)}-\frac{\gamma_0}{b_0g}\right)\right\} \ = \nonumber \\[2.0ex]
& =\  \hat c(\mu_{\rm pt})\ U(\mu_{\rm pt},\mu_{\rm had}) \ .
\end{align}
The second factor on the rhs is known non-perturbatively as a product of continuum
SSFs $\sigma(u)$; cf. \req{eq:evol}. The first factor can be safely computed at next-to-leading 
order (NLO) in perturbation theory, provided the scale $\mu_{\rm pt}$ is high enough 
to render NNLO effects negligible. The full procedure for the construction of  
$U(\mu_{\rm pt}, \mu_{\rm had})$ has been introduced in \cite{Capitani:1998mq} for
the running quark mass in the quenched approximation, and subsequently
applied in several contexts  (for a recent review, see \cite{Sommer:2006sj}). The reader
is referred to the above-mentioned works for a detailed description of the method. More
specifically, since the present work concerns two-flavour QCD, we follow closely the work 
of ref.~\cite{DellaMorte:2005kg} on the running quark mass with $N_{\rm f}=2$ dynamical quarks.

Concerning the choice of $[\Gamma_{\rm A},\Gamma_{\rm B},\Gamma_{\rm C}]$ 
and $\alpha$, we observe that
in our quenched studies \cite{Guagnelli:2005zc,Palombi:2007dr} we 
have considered five possible non-trivial Dirac structures that preserve
cubic symmetry at vanishing external momenta, i.e. 
\begin{align}
[\Gamma_{\rm A},\Gamma_{\rm B},\Gamma_{\rm C}] = \left\{ \ \right. & [\gamma_5,\gamma_5,\gamma_5],\ \epsilon_{ijk}[\gamma_i,\gamma_j,\gamma_k],\ [\gamma_5,\gamma_k,\gamma_k], \nonumber \\[1.2ex]
& [\gamma_k,\gamma_5,\gamma_k],\ [\gamma_k,\gamma_k,\gamma_5]\ \left . \right\} \ 
\end{align}
(where a sum over repeated indices is understood), and various possible values of the 
$\alpha$ parameter, namely $\alpha=\{0,1/2\}$ for the static operators and $\alpha=\{0,1,3/2\}$
for the relativistic ones. Not all of the resulting renormalisation 
schemes were equally well suited to our purposes: some of 
them were characterised by a RG running with a slow perturbative convergence at
NLO; cf. refs.~\cite{Palombi:2005zd,Palombi:2006pu}.
This rendered the matching of perturbative and non-perturbative
running at $\mu_{\rm pt}$ (cf. \req{RGrunexp}) less reliable and the systematics 
hard to control; see sect.~3.3 below for more details. The same considerations 
are valid in the present case of two dynamical quarks. For the sake of brevity 
we will concentrate only in those schemes which have been found to be best 
behaved in the present unquenched study. These optimal schemes are specified in 
Table~\ref{tab:table1}. A complete account of our results in all schemes 
considered is available upon request.

\begin{table}[!t]
\begin{center}
\begin{tabular}{c|cccc}
\Hline \\[-10pt]
$Q$ & \ \ $\Gamma_{\rm A}$ \ \ & \ \ $\Gamma_{\rm B}$ \ \ & \ \ $\Gamma_{\rm C}$ \ \ & \ \ $\alpha$ \ \  \\
\hline \\[-5pt]
$Q_1^+$  & $\gamma_5$ & $\gamma_5$ & $\gamma_5$ & $0$ \\[0.7ex] 
$Q_1^-$  & $\gamma_k$ & $\gamma_5$ & $\gamma_k$ & $1$ \\[1.5ex]
${\cQ'}_1^{+}$ & $\gamma_5$ & $\gamma_5$ & $\gamma_5$ & $1/2$ \\[0.7ex]
${\cQ'}_2^{+}$ & $\gamma_5$ & $\gamma_5$ & $\gamma_5$ & $0$ \\[0.7ex] 
${\cQ'}_3^{+}$ & $\gamma_5$ & $\gamma_5$ & $\gamma_5$ & $0$ \\[0.7ex] 
${\cQ'}_4^{+}$ & $\gamma_5$ & $\gamma_5$ & $\gamma_5$ & $0$ \\[1.0ex] 
\Hline
\end{tabular}
\end{center}
\caption{\small Optimal renormalisation schemes for the various four-quark operators.}
\label{tab:table1}
\end{table}

The non-universal two-loop coefficients of the anomalous dimensions
Eq.~(\ref{anomalousdimen}) for the operators of interest in the aforementioned optimal schemes 
read \cite{Palombi:2005zd,Palombi:2007dr}
\begin{alignat}{3}
\gamma_1^+     & = \phantom{-}\frac{1}{(4\pi)^2}\bigl[\phantom{-}0.0828(48) +\ 0.03200(28) N_{\rm f}\bigr]\ , \\[0.5ex]
\gamma_1^-     & = -\frac{1}{(4\pi)^2}\bigl[-0.6880(24) +\ 0.12648(16) N_{\rm f}\bigr]\ , \\[0.5ex]
\gamma_1^{(1)} & = -\frac{1}{(4\pi)^2}\bigl[\phantom{-}1.345(2)\phantom{0}+\ 0.0008(2)N_{\rm f}\bigr]\ , \\[0.5ex]
\gamma_1^{(2)} & = -\frac{1}{(4\pi)^2}\bigl[-1.251(1)\phantom{0} +\ 0.11637(8)N_{\rm f}\bigr]\ ,  \\[0.5ex]
\gamma_1^{(3)} & = -\frac{1}{(4\pi)^2}\bigl[-0.327(3)\phantom{0} +\ 0.1211(2) N_{\rm f}\bigr]\ ,  \\[0.5ex]
\gamma_1^{(4)} & = -\frac{1}{(4\pi)^2}\bigl[-0.146(1)\phantom{0} +\ 0.06784(8) N_{\rm f}\bigr]\ .
\end{alignat}

\section{Non-perturbative computation of the RG running}

\subsection{Simulations details}

Our simulations are based on the regularisation of relativistic quarks by
the non-perturbatively $\Oa$ improved Wilson action, with the Sheikoleslami-Wohlert 
(SW) coefficient $c_{\rm sw}$ determined in~\cite{Jansen:1998mx}. 
Static quarks have been instead discretised as proposed in \cite{DellaMorte:2005yc}. 
In particular, all results reported in this work refer to the so-called 
HYP2 action, i.e. the lattice static action of \cite{Eichten:1989zv}, 
with the standard parallel transporter $U(0,x)$ replaced by the temporal hypercubic 
link introduced in \cite{Hasenfratz:2001hp}, and a choice of the smearing parameters 
 $(\alpha_1,\alpha_2,\alpha_3) = (1.0,1.0,0.5)$. The latter minimises the 
quenched static self-energy, providing the largest exponential increase of the
signal-to-noise ratio in the static-quark propagator, when compared to the original Eichten-Hill 
action. The minimum  of the static self-energy is shifted by internal quark-loops only at NLO in perturbation theory: such shift is thus expected to be relatively small. 

With the above prescriptions, the SSFs have been computed at six different values 
of the SF renormalised coupling, corresponding to six different physical lattice lengths 
$L$. For each physical volume three different values of the lattice spacing have been simulated, corresponding to lattices with $L/a=6,8,12$ (and $2L/a=12,16,24$ respectively)
for the computation of $Z(g_0,a/L)$ (and $Z\left(g_0,a/(2L)\right)$). 

The gauge configuration ensemble used in the present work (generated with
$N_{\rm f}=2$
dynamical fermions) and the tuning of the lattice parameters
$(\beta,\kappa)$ have
been taken over from \cite{DellaMorte:2005kg}. All technical details 
concerning these
dynamical fermion simulations are discussed in that work. The one
technical aspect that makes
a significant difference in our case concerns the perturbative value of
the boundary improvement
coefficient $c_{\rm t}$ \cite{Luscher:1992an}. As pointed out in
\cite{DellaMorte:2005kg}, the gauge configurations at the three weakest
couplings have been produced using the one--loop perturbative
estimate of $c_{\rm t}$ \cite{Luscher:1992an}, except for ($L/a~=~6$,\
$\beta = 7.5420$) and
($L/a = 8,\ \beta = 7.7206$). For these two cases and for the three stronger
couplings, the
two--loop value of $c_{\rm t}$ \cite{Bode:1999sm} has been used. We have
enforced the same
$c_{\rm t}$ values in the valence propagators. Comparison of the results
of two different simulations, namely
\begin{alignat}{5}
& \bar g_{\rm\scriptscriptstyle SF}^2 = 1.5031(25) \ , \quad & L/a=6\ ,
\quad & \beta = 7.5000\ , \quad & \kappa = 0.1338150 \ , \quad & c_{\rm t}
= {\rm one-loop}\ ,  \nonumber \\[0.0ex]
& \bar g_{\rm\scriptscriptstyle SF}^2 = 1.5078(44) \ , \quad & L/a=6\ ,
\quad & \beta = 7.5420\ , \quad & \kappa = 0.1337050 \ , \quad & c_{\rm t}
= {\rm two-loop}\ .  \nonumber
\end{alignat}
shows that the renormalisation factor $Z_P$ of the pseudoscalar density,
analysed in
\cite{DellaMorte:2005kg}, is subject to a relative 4 per mille variation,
corresponding to a mild discrepancy of about $1.5\sigma$ with regards to our
statistical uncertainty. Unfortunately, for the four-fermion
correlation functions this remains true only for the operator $Q_1^-$,
while the $Q_1^+$ and ${\cQ'}_k^+$ cases show relative variations
of the order of 1-2\%, corresponding to differences of about
$2.5\sigma$ with regards to the statistical precision. We expect that, for
a given renormalised coupling
$\gbar^2_{\rm\scriptscriptstyle SF}$, this effect diminishes at finer lattice resolutions $a/L$ 
(i.e. closer to the continuum), while it becomes more pronounced in the strong coupling region, 
at constant $L/a$. In principle, this problem can be removed by performing all simulations 
with a two--loop estimate of $c_{\rm t}$ and/or a smaller resolution $a/L$. 
As further dynamical simulations are beyond the scope of the present work,
we limit ourselves in stating that our results are subject to this
ill-controlled systematic uncertainty, which, in view of the fact that 
the one-loop value of $c_{\rm t}$ is only used in the weak coupling region,
is however not expected to be significant for our final results.
In this respect, we have checked that the (final) overall result is unaffected when either
$c_{\rm t}$ is employed at this coupling. It is also encouraging that,
as we will see below, including or discarding the
$L/a = 6$ data-points in the continuum extrapolations does not alter the 
final results significantly.

Numerical results are collected in Tables~\ref{tab:tab2}--\ref{tab:tab4}.
Statistical errors were computed by a jackknife analysis.
The estimates of the autocorrelation times, calculated with the     
autocorrelation function method, the method of ref.~\cite{Wolff:2003sm} and the      
binning method, were found to be compatible.

\subsection{Continuum extrapolation of the step-scaling functions}

\begin{figure}[!t]
  \begin{center}
    \vskip 1.5cm
    \epsfig{figure=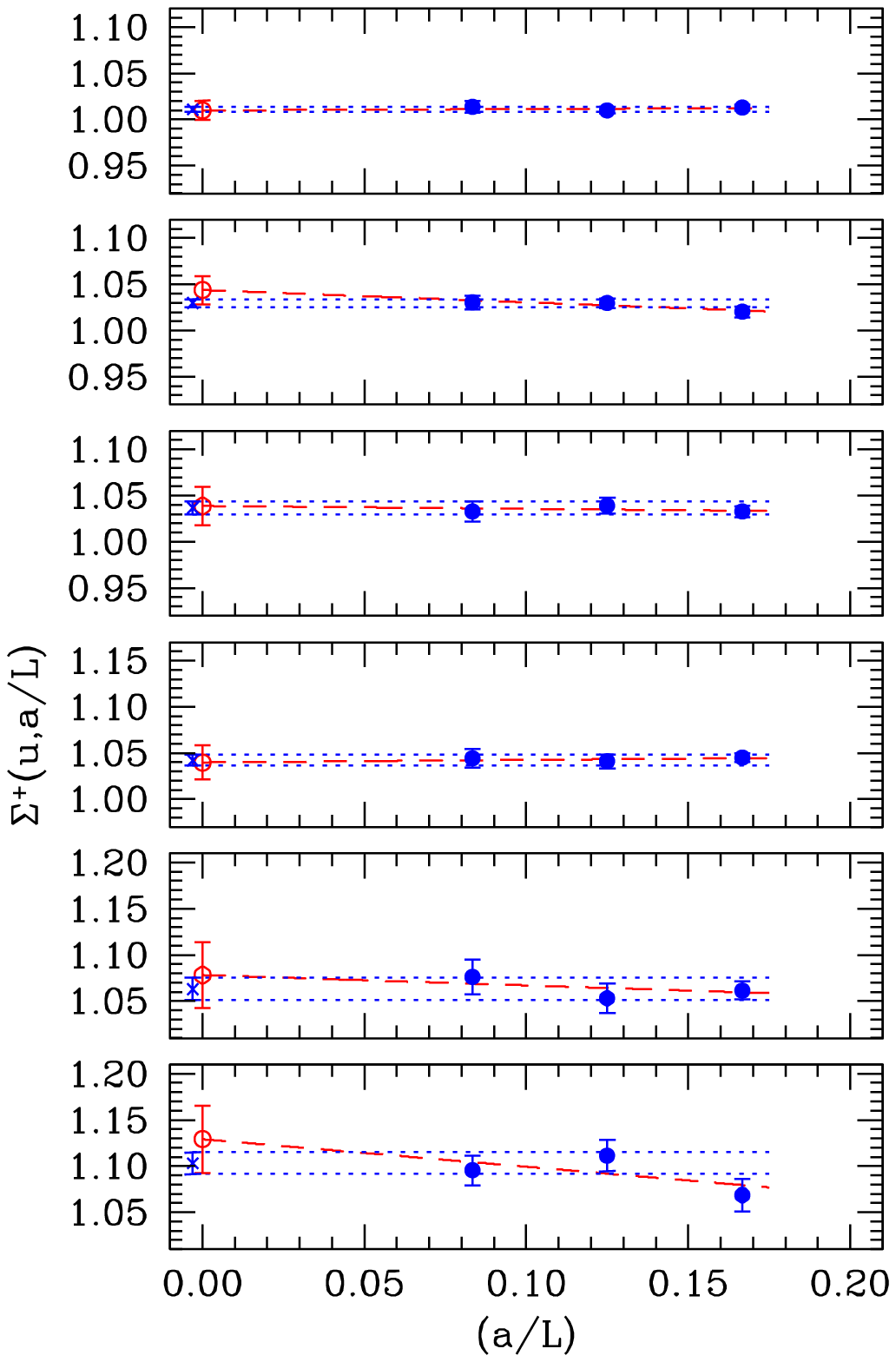, width=6.5 true cm}
    \epsfig{figure=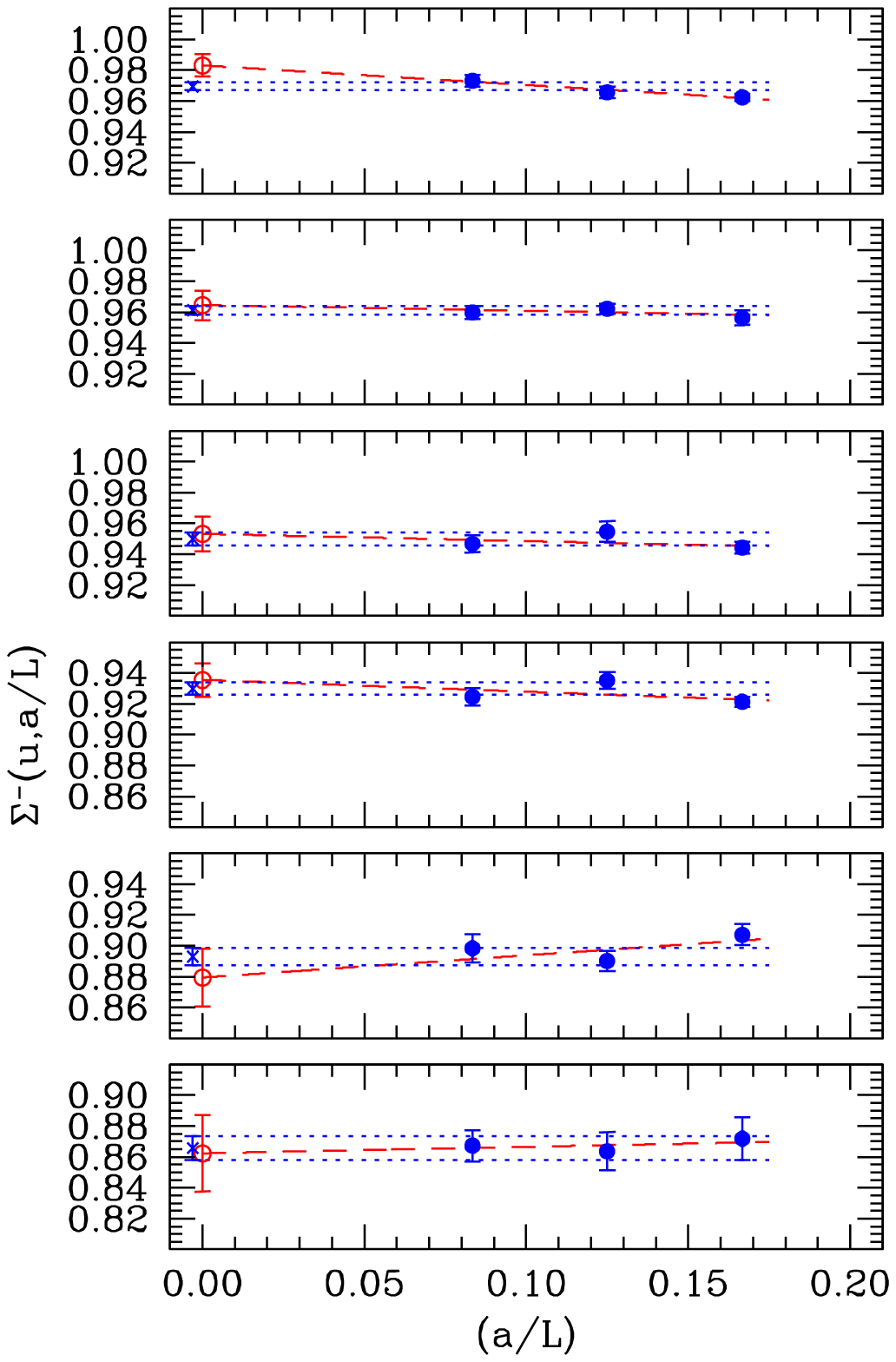, width=6.5 true cm}
  \end{center}
  \vskip -0.4cm
  \caption{\small Continuum extrapolation of the SSFs for $Q_1^+$ (left) and $Q_1^{-}$ (right). 
    The renormalised coupling increases from top to bottom. Blue dotted lines and the 
    blue cross at $a/L=0$ correspond to weighted averages of the $L/a=8,12$ data, red dashed 
    lines and the red $a/L=0$ open point to linear extrapolations in $a/L$ of the three data.}
  \label{fig:extrap1}
\end{figure}

\begin{figure}[!h]
  \begin{center}
    \vskip 1.5cm
    \epsfig{figure=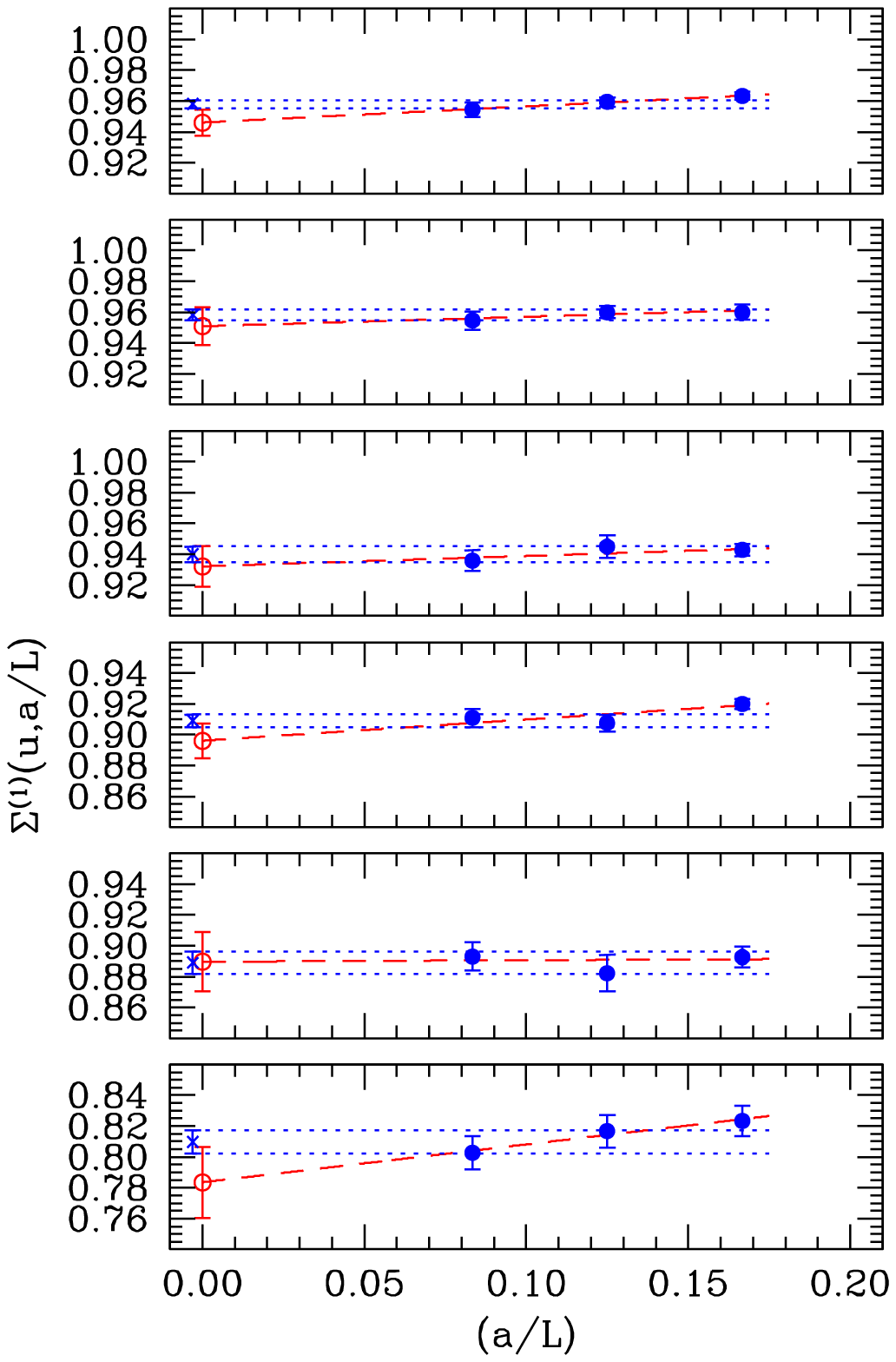, width=6.5 true cm}
    \epsfig{figure=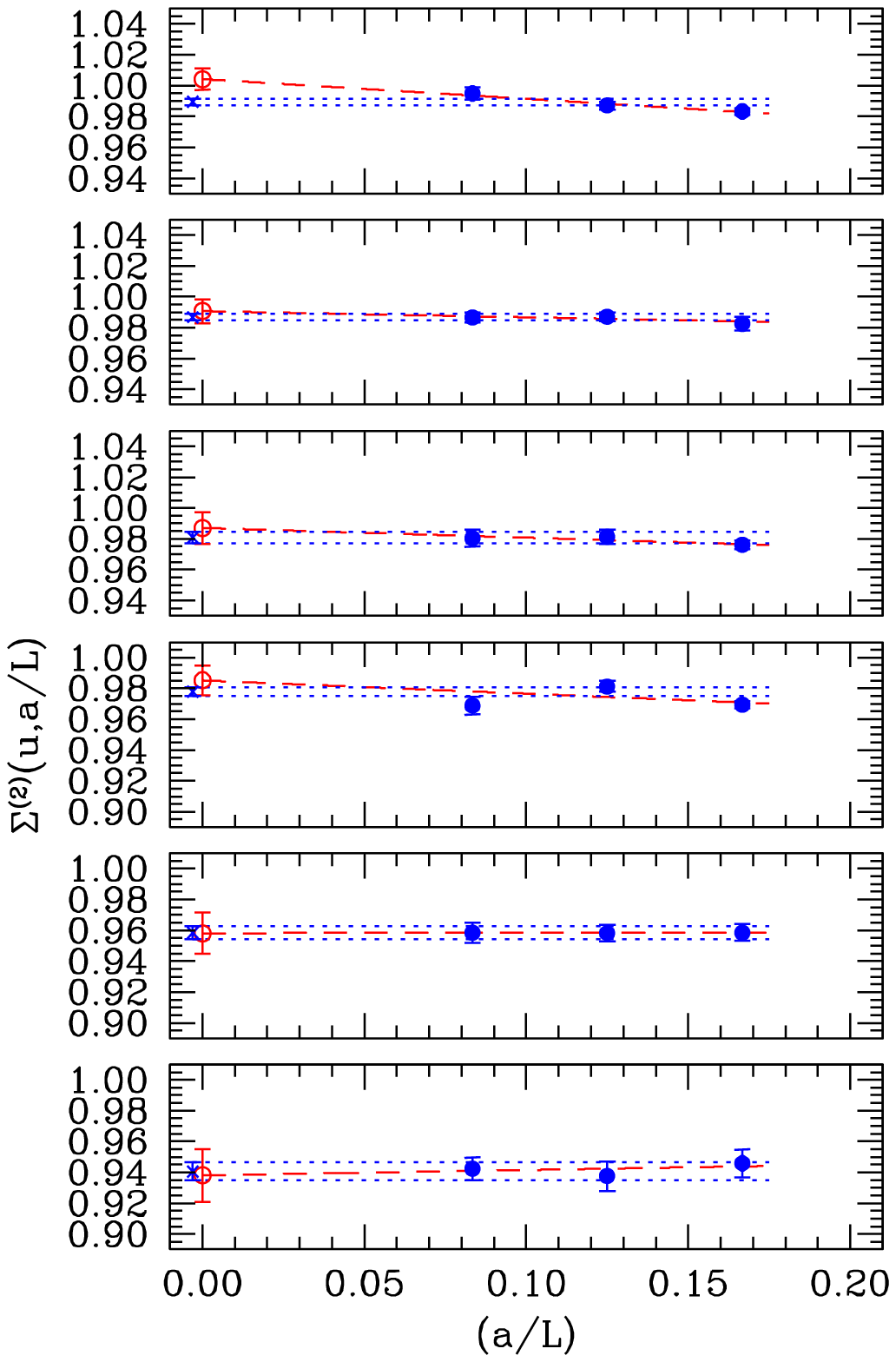, width=6.5 true cm}
       \end{center}
  \vskip -0.4cm
  \caption{\small Continuum extrapolation of the SSFs for ${\cQ'}_1^{+}$ (left) and ${\cQ'}_2^{+}$ (right). 
    The renormalised coupling increases from top to bottom. Blue dotted lines and the 
    blue cross at $a/L=0$ correspond to weighted averages of the $L/a=8,12$ data, red dashed 
    lines and the red $a/L=0$ open point to linear extrapolations in $a/L$ of the three data.}
  \label{fig:extrap2}
\end{figure}

\begin{figure}[!h]
  \begin{center}
    \vskip 1.5cm
    \epsfig{figure=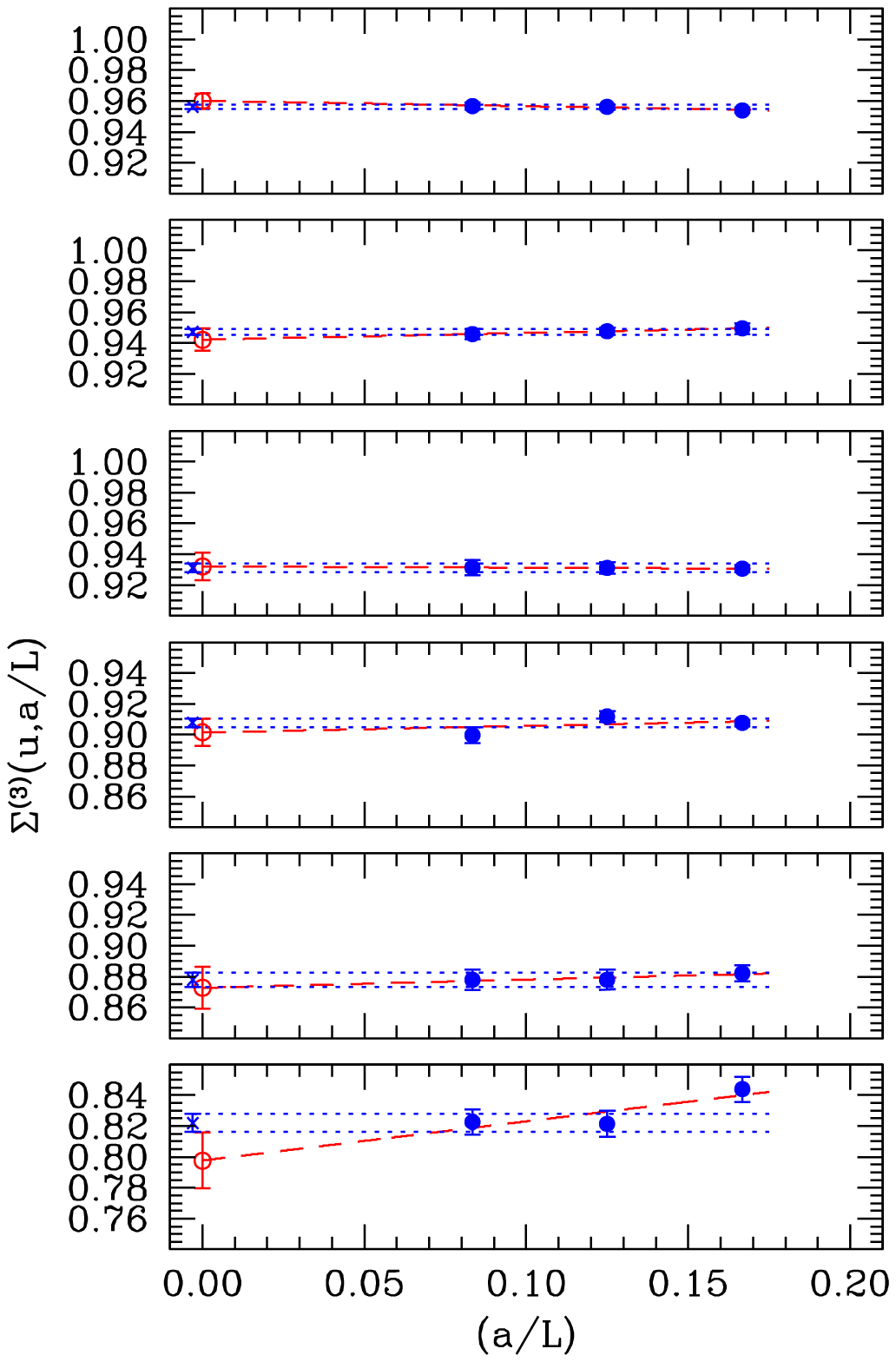, width=6.5 true cm}
    \epsfig{figure=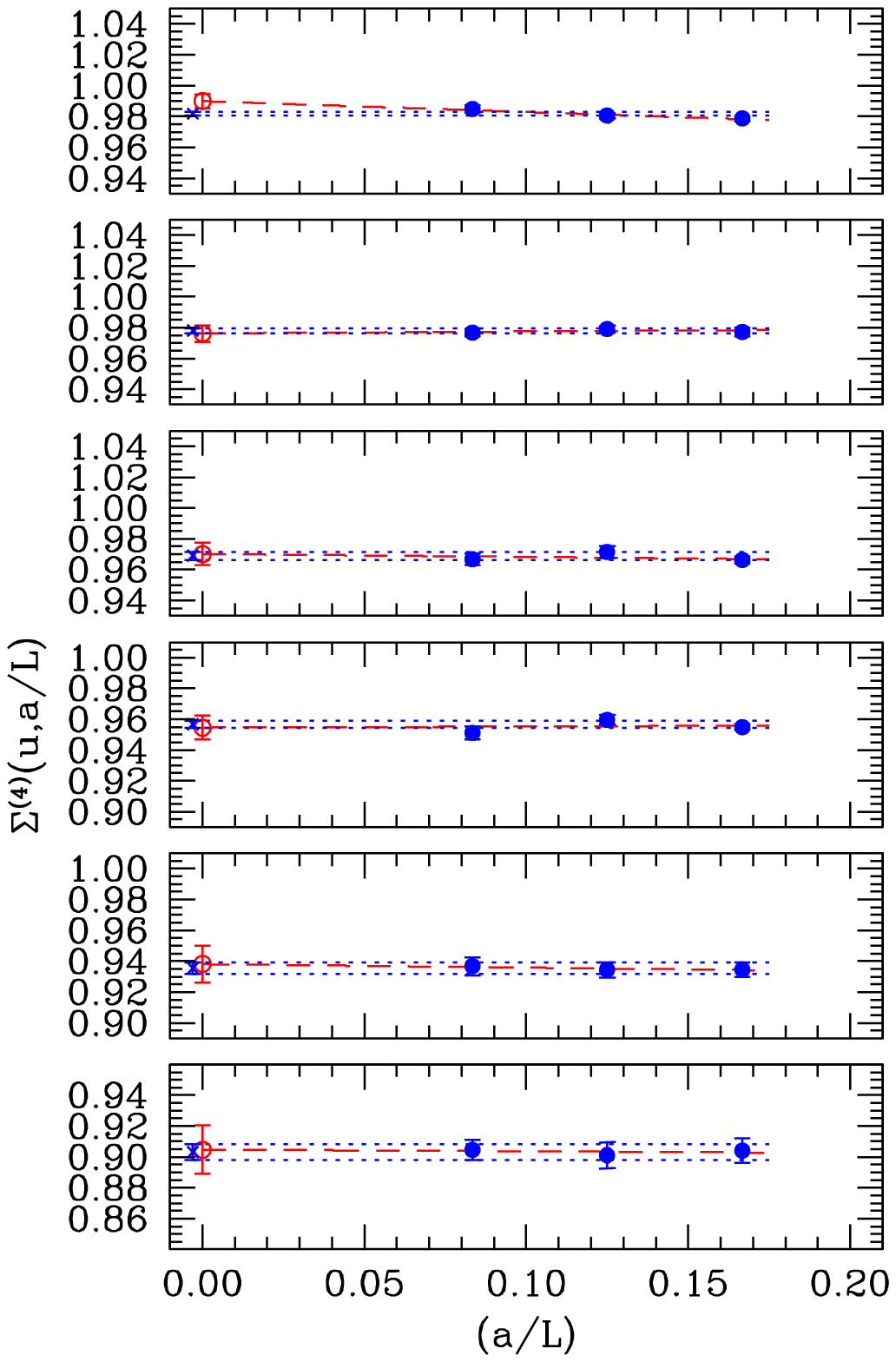, width=6.5 true cm}
  \end{center}
  \vskip -0.4cm
  \caption{\small Continuum extrapolation of the SSFs for ${\cQ'}_3^{+}$ (left) and ${\cQ'}_4^{+}$ (right). 
    The renormalised coupling increases from top to bottom. Blue dotted lines and the 
    blue cross at $a/L=0$ correspond to weighted averages of the $L/a=8,12$ data, red dashed
    lines and the red $a/L=0$ open point to linear extrapolations in $a/L$ of the three data.}
  \label{fig:extrap3}
\end{figure}

Since we do not implement $\Oa$ improvement of four-fermion operators,
the only linear cutoff effects that are removed from $\Sigma(u,a/L)$
are those cancelled by the SW term in the fermion action. Therefore, we
expect SSFs to approach the continuum limit linearly in $a/L$ and correspondingly we
fit to the ansatz
\begin{equation}
\Sigma(u,a/L) = \sigma(u) + \rho(u)(a/L)\ .
\end{equation}
In practice it is often observed that the data corresponding
to $L/a=8,12$ are compatible within errors, whereas the $L/a=6$ result,
bearing the largest cutoff effects, is off. This suggests that,
in analogy to~\cite{DellaMorte:2005kg}, a weighted average of the two
finest lattice results may be a reliable estimate of the continuum limit 
value. We have checked that, in most cases, linear fits to all three 
data-points and weighted averages of the two results from the finer 
lattices lead to continuum limit estimates, compatible within one standard 
deviation; cf. Figs.~\ref{fig:extrap1}, \ref{fig:extrap2} and \ref{fig:extrap3}. 
Fit results are reported in Table~\ref{tab:contsig}. Since the discretisation 
errors are ${\rm O}(a)$ and not ${\rm O}(a^2)$ as in \cite{DellaMorte:2005kg}, 
we conservatively quote, as our best results, those obtained from linear 
extrapolations involving all three data-points.

It should be added that, besides the HYP2 action, we have also tried other 
static quark action varieties, namely the APE and the 
HYP1 ones (see ref.~\cite{DellaMorte:2005yc}), which differ from HYP2 by 
${\rm O}(a^2)$ lattice artefacts. Since the static four-quark operators are 
not $\Oa$ improved, it is reasonable to expect significant discretisation 
effects at the coarsest lattice spacing, which would enable combined fits 
of the data from all three actions, constrained to a common value in the 
continuum limit. Unsurprisingly, the situation turned out to be similar to that
of~\cite{Palombi:2007dr}, in that data obtained with the above actions do not differ 
noticeably (even at L/a = 6) and are very strongly correlated. Consequently,
a combined continuum extrapolation affects the continuum limit results only marginally, with
the relative error decreasing only by a few percent. For this reason we only
quote results from the HYP2 analysis. 
\begin{table}[t]
  \footnotesize
  \centering
  \renewcommand{\arraystretch}{1.25}
  \begin{tabular}{ccccccc} 
    \Hline\\[-3.0ex]
    $u$ & ~~{$\sigma^{+}(u)$}  ~~
        & ~~{$\sigma^{-}(u)$}  ~~
        & ~~{$\sigma^{(1)}(u)$}~~
        & ~~{$\sigma^{(2)}(u)$}~~
        & ~~{$\sigma^{(3)}(u)$}~~
        & ~~{$\sigma^{(4)}(u)$}~~ \\[0.0ex] \hline\\[-3.0ex]
    0.9793 & 1.010(11)  & 0.983(07)  & 0.946(08) & 1.004(07) & 0.960(05) & 0.990(05) \\
    1.1814 & 1.044(15)  & 0.965(10)  & 0.951(12) & 0.991(08) & 0.942(07) & 0.976(05) \\
    1.5078 & 1.039(21)  & 0.953(11)  & 0.932(13) & 0.987(10) & 0.932(09) & 0.970(07) \\
    2.0142 & 1.040(18)  & 0.936(11)  & 0.896(11) & 0.985(10) & 0.901(09) & 0.955(08) \\
    2.4792 & 1.078(35)  & 0.879(19)  & 0.890(19) & 0.958(13) & 0.873(14) & 0.938(12) \\
    3.3340 & 1.129(37)  & 0.862(25)  & 0.784(23) & 0.938(17) & 0.798(18) & 0.905(16) \\[0.5ex]
    \Hline
  \end{tabular}
  \caption{
    \small Results of the continuum limit extrapolation of the lattice step-scaling
    functions $\Sigma^\pm$ and $\Sigma^{(k)}$. Data have been fitted from all available 
    lattice resolutions as linear functions in $(a/L)$.
  }\label{tab:contsig}
\end{table}

\subsection{RG running in the continuum limit}

\begin{figure}[!h]
  \vskip -1.2cm
   \begin{center}
     \vskip 1.5cm
     \hskip -0.2cm \epsfig{figure=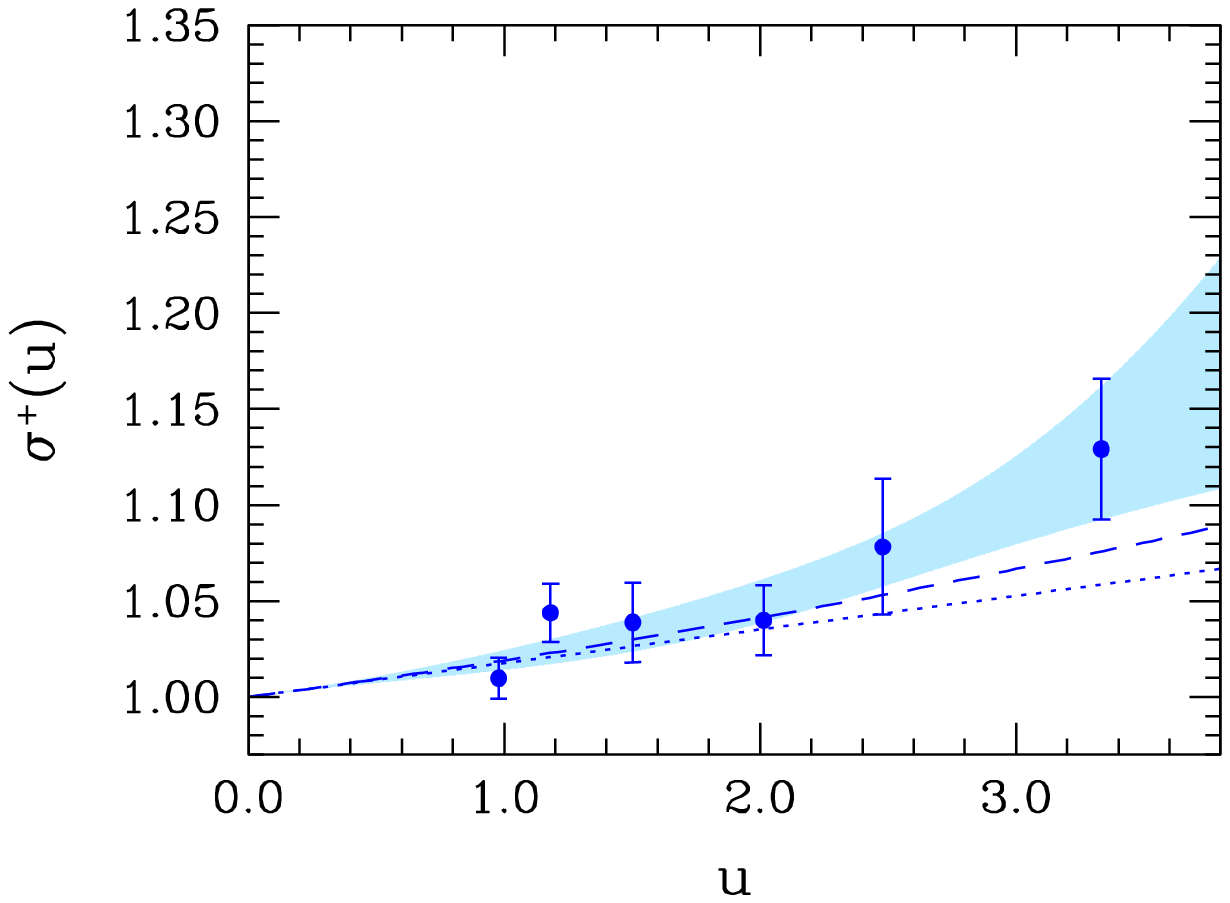, width=6.5 true cm}
     \epsfig{figure=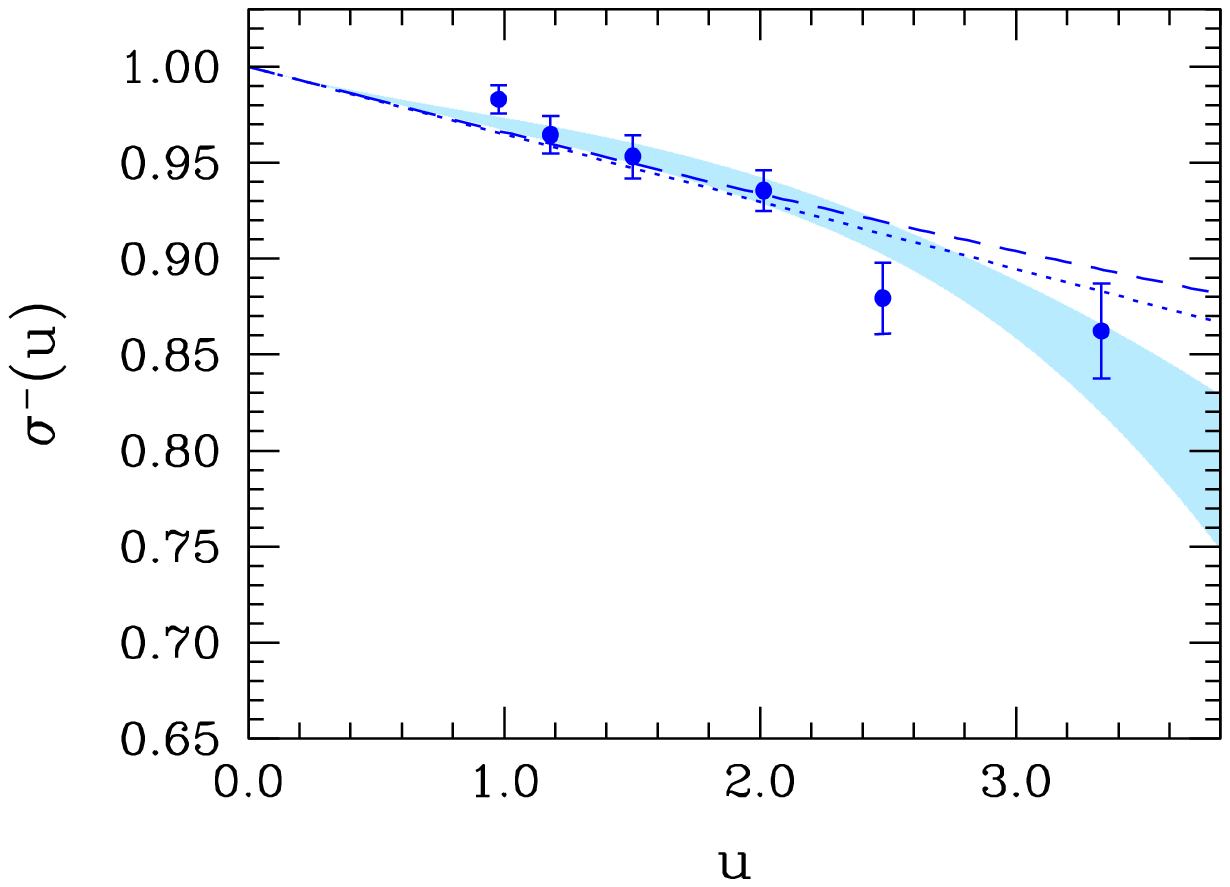, width=6.5 true cm}
   \end{center}
   \vskip -0.4cm
   \caption{\small 
     The step-scaling functions $\sigma^+$ and $\sigma^-$ (discrete points) as obtained 
     non-perturbatively. The shaded area is the one sigma band obtained by fitting the points 
     to a polynomial as discussed in the text. The dotted (dashed) line is the LO (NLO) perturbative result.}
   \label{fig:sigma1}
 \end{figure}

\begin{figure}[!hb]
  \vskip -1.2cm
  \begin{center}
    \vskip 1.5cm
    \hskip -0.2cm \epsfig{figure=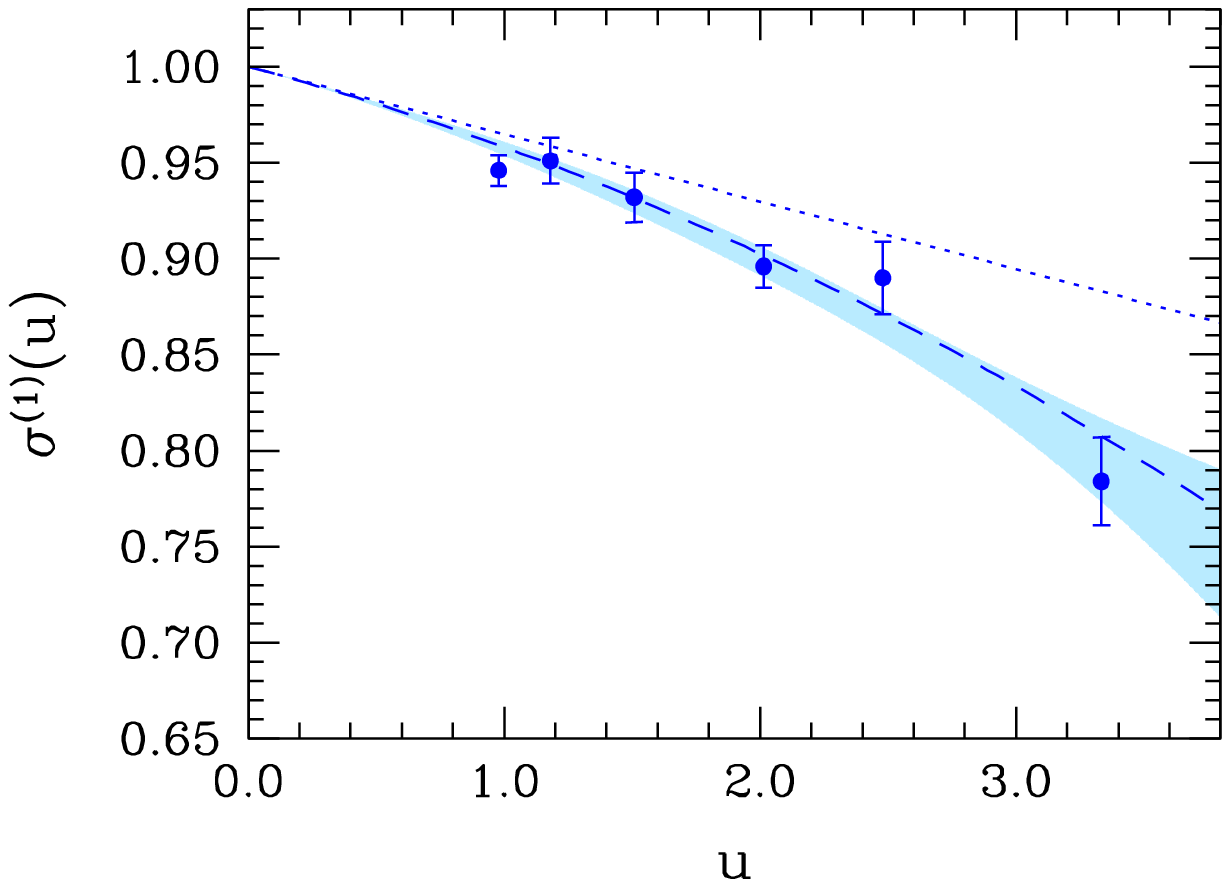, width=6.5 true cm}
    \epsfig{figure=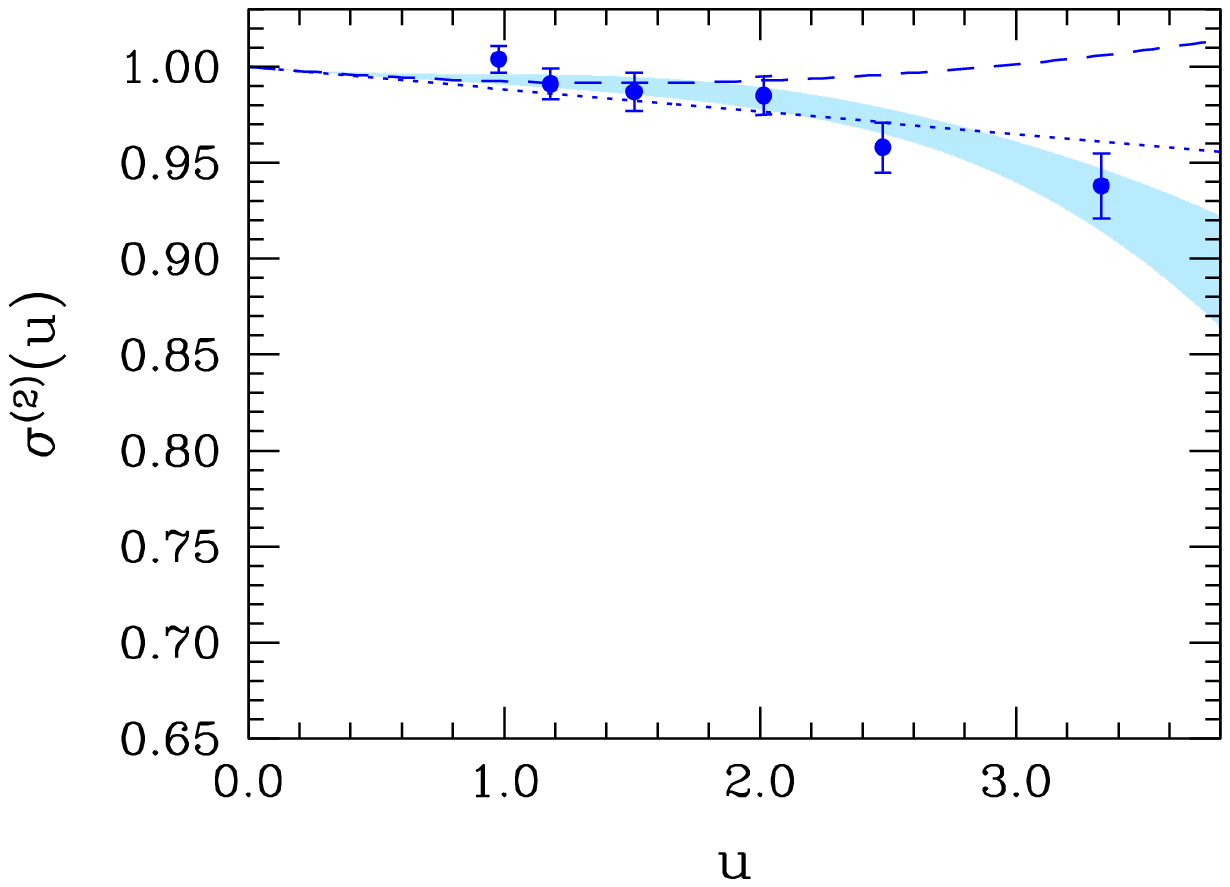, width=6.5 true cm}
  \end{center}
  \vskip -0.4cm
  \caption{\small 
     The step-scaling functions $\sigma^{(1)}$ and $\sigma^{(2)}$ (discrete points) as obtained 
     non-perturbatively. The shaded area is the one sigma band obtained by fitting the points 
     to a polynomial as discussed in the text. The dotted (dashed) line is the LO (NLO) perturbative result.
     }
  \label{fig:sigma2}
\end{figure}

\begin{figure}[!hb]
  \vskip -1.2cm
  \begin{center}
    \vskip 1.5cm
     \hskip -0.2cm \epsfig{figure=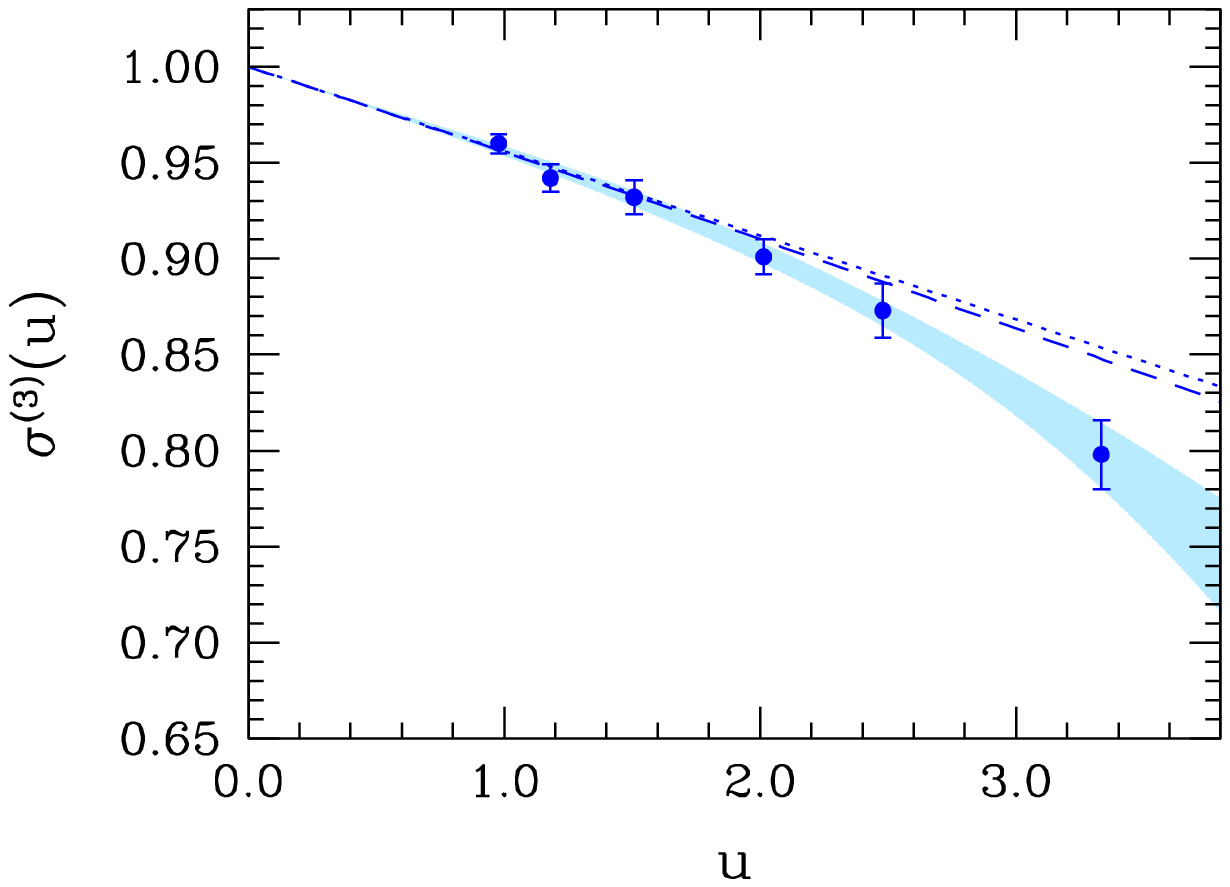, width=6.5 true cm}
     \epsfig{figure=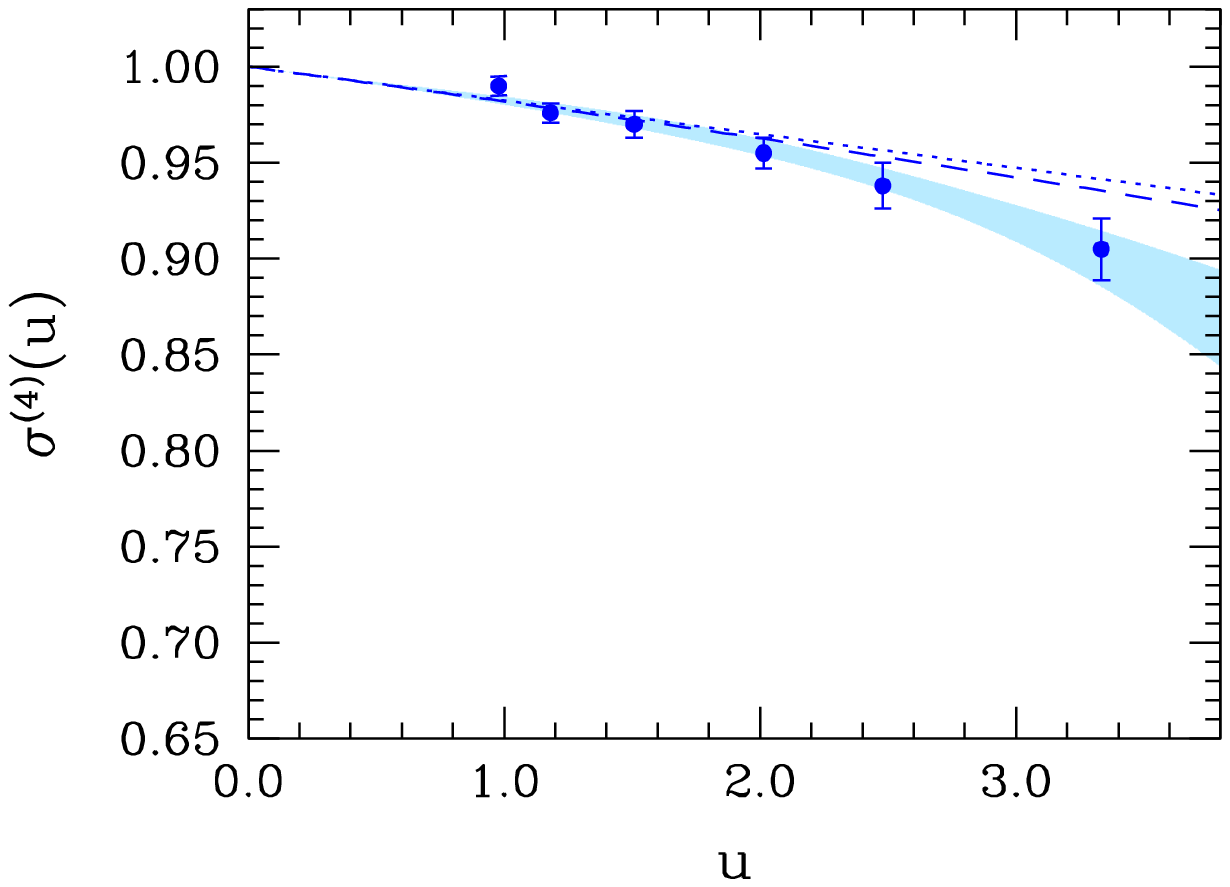, width=6.5 true cm}
  \end{center}
  \vskip -0.4cm
  \caption{\small 
     The step-scaling functions $\sigma^{(3)}$ and $\sigma^{(4)}$ (discrete points) as obtained 
     non-perturbatively. The shaded area is the one sigma band obtained by fitting the points 
     to a polynomial as discussed in the text. The dotted (dashed) line is the LO (NLO) perturbative result.
     }
  \label{fig:sigma3}
\end{figure}

In order to compute the RG running of the operators in the
continuum limit as described in \cite{DellaMorte:2005kg},
the continuum SSFs have to be fitted to some functional 
form. The simplest choice is represented by a polynomial
\begin{equation}
\sigma(u)=1+s_1u+s_2u^2+s_3u^3 + \dots \ ,
\end{equation}
whose form is motivated by the perturbative series, with coefficients
\begin{gather}
\label{eq:pt_s1}
s_1 = {\gamma}_0\ln 2 \ , \\[1.5ex]
\label{eq:pt_s2}
s_2 = {\gamma}_1\ln 2 + \left[\frac{1}{2}(\gamma_0)^2 + b_0\gamma_0\right](\ln 2)^2 \ .
\end{gather}
It is worth stressing that $s_1$ is universal and independent of $N_{\rm f}$, 
whereas $s_2$ carries a dependence upon $N_{\rm f}$ via $b_0$ and $\gamma_1$,
with the latter coefficient introducing a scheme dependence. In our fits we truncated the polynomial
at ${\rm O}(u^3)$. The fits have been performed with $s_1$ fixed to its perturbative value 
and $s_2$, $s_3$ left as free parameters. Fit results are shown in Figs.~\ref{fig:sigma1}, \ref{fig:sigma2} and 
\ref{fig:sigma3}. Fitted values of $s_2$ turned  out to be close to the perturbative 
prediction of Eq.~(\ref{eq:pt_s2}), with the exception of ${\cal Q'}_2^+$.

\begin{figure}[!hb]
  \vskip -1.4cm
  \begin{center}
    \vskip 1.5cm
     \hskip -0.2cm \epsfig{figure=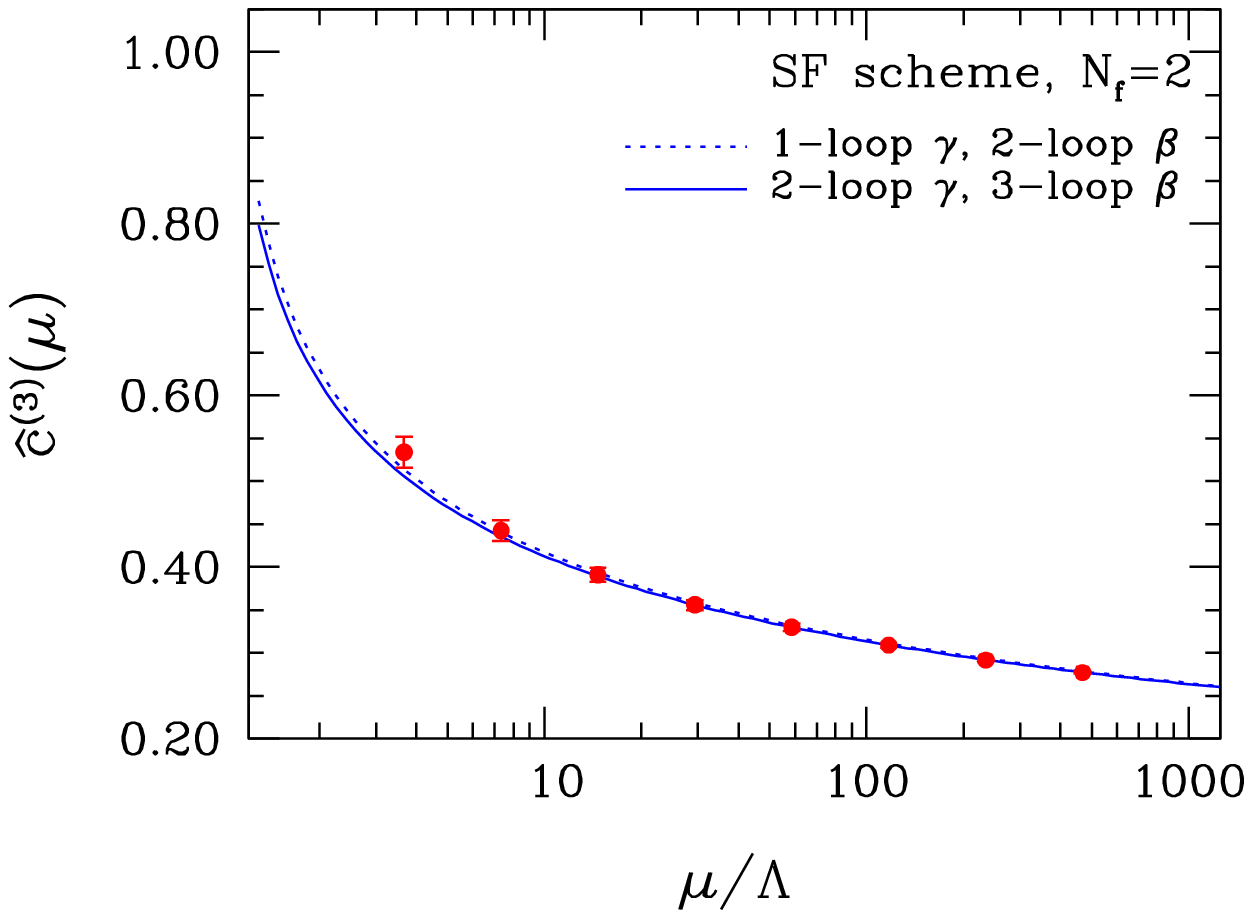, width=6.5 true cm}
     \epsfig{figure=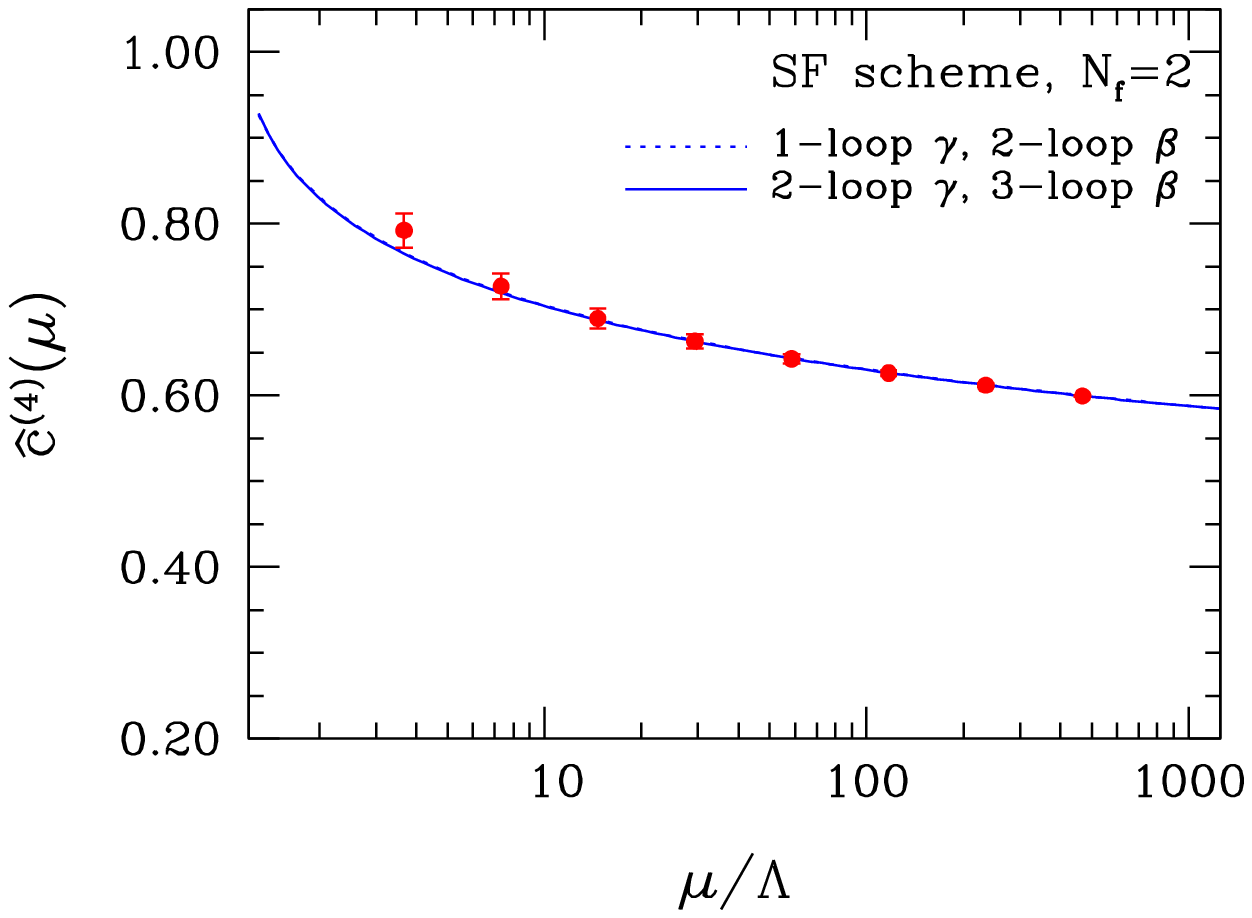, width=6.5 true cm}
  \end{center}
  \vskip -0.3cm
  \caption{\small
    RG-running of of the four-quark operators obtained non-perturbatively 
    (discrete points) at specific values of the renormalisation scale $\mu$, 
    in units of $\Lambda$. The lines are perturbative results at the order 
    shown for the Callan-Symanzik $\beta$ function and the operator 
    anomalous dimension $\gamma$.
  }
  \label{fig:RGrunning}
\end{figure}

Once the continuous SSFs have been obtained as functions of the  
renormalised coupling, the ratios $\hat c$ (cf. \req{eq:rgi}) are obtained recursively.
The low-energy scale $\mu_{\rm had}=\lmax^{-1}$ is implicitly defined in this work
through the condition $\gbar^2_{\rm\scriptscriptstyle SF}(\lmax) = 4.61$, as explained in \cite{DellaMorte:2005kg}.
This scale is chosen so that the renormalisation constants
$Z(g_0,a\mu_{\rm had})$ can be computed in the accessible $g_0$-range commonly used in large-volume simulations.
The non-perturbative RG running of the six operators of interest
are shown in Fig.~\ref{fig:RGrunning}.

\begin{table}[!t]
  \footnotesize
  \centering
  \renewcommand{\arraystretch}{1.25}
  \begin{tabular}{cccccccc}
    \Hline\\[-3.0ex]
    $n$ & $u$ & $\hat c^{\,+}(L_{\rm max}^{-1})$ & $\hat c^{\,-}(L_{\rm max}^{-1})$ & $\hat c^{\,(1)}(L_{\rm max}^{-1})$ & $\hat c^{\,(2)}(L_{\rm max}^{-1})$ & $\hat c^{\,(3)}(L_{\rm max}^{-1})$ & $\hat c^{\,(4)}(L_{\rm max}^{-1})$ \\ \\[-3.0ex] \hline\\[-3.0ex]
    0 & 4.610 & 1.246      & 0.551      & 0.807      & 0.680      & 0.524      & 0.776   \\
    1 & 3.032 & 1.225(26)  & 0.564(10)  & 0.775(14)  & 0.732(09)  & 0.532(08)  & 0.788(09)  \\
    2 & 2.341 & 1.212(38)  & 0.566(14)  & 0.773(20)  & 0.751(12)  & 0.538(11)  & 0.794(12)  \\
    3 & 1.918 & 1.205(46)  & 0.564(16)  & 0.773(24)  & 0.759(15)  & 0.541(13)  & 0.797(14)  \\
    4 & 1.628 & 1.202(53)  & 0.561(18)  & 0.772(27)  & 0.762(18)  & 0.541(14)  & 0.797(16)  \\
    5 & 1.414 & 1.201(60)  & 0.558(20)  & 0.772(30)  & 0.763(20)  & 0.541(15)  & 0.797(18)  \\
    6 & 1.251 & 1.201(66)  & 0.554(21)  & 0.771(33)  & 0.763(22)  & 0.540(17)  & 0.797(19)  \\
    7 & 1.121 & 1.202(71)  & 0.551(22)  & 0.770(35)  & 0.763(24)  & 0.539(18)  & 0.796(20)  \\
    8 & 1.017 & 1.202(76)  & 0.548(24)  & 0.770(37)  & 0.762(26)  & 0.538(19)  & 0.795(22)  \\ \\[-3.0ex] \Hline
  \end{tabular}
  \caption{\small
    Perturbative matching (cf. Eq.~(\ref{RGrunexp})) for various choices of the matching scale $\mu_{\rm pt} = 2^n\mu_{\rm had}$.
  }\label{tab:phiratio_res}
\end{table}

As discussed in our former quenched study~\cite{Palombi:2007dr}, the 
main criterion for selecting robust schemes amounts to checking that 
the systematic uncertainty present in our final results, due to the NLO truncation of the
perturbative matching at the scale $\mu_{\rm pt}\equiv 2^n\mu_{\rm had}$, is well under 
control. This in turn requires an estimate of the size of the NNLO
contribution to $\hat{c}$. To this purpose we have re-computed 
${\hat{c}}$ with two different ans\"atze for the NNLO anomalous dimensions 
${\gamma}_2$: (i) we set $|{\gamma}_2/{\gamma}_1|=|{\gamma}_1/{\gamma}_0|$;
(ii) we perform a two-parameter fit to the SSF with $s_1$,$s_2$ fixed to their perturbative
values and $s_3$,$s_4$ left as free parameters, and then estimate ${\gamma}_2$ by
equating the resulting value of $s_3$ to its perturbative expression
\begin{align}
s_3 = \gamma_2\ln 2 &+ \left[\gamma_0\gamma_1 + 2b_0\gamma_1 + b_1\gamma_0\right](\ln 2)^2 + \nonumber \\[2.0ex] 
& + \left[\frac{1}{6}\gamma_0^3 + b_0 \gamma_0^2+\frac{4}{3}b_0^2\gamma_0\right](\ln 2)^3 \ .
\end{align}
The optimal schemes specified in Table~\ref{tab:table1} are precisely those for which
the aforementioned determinations of the effective $\gamma_2$ lead to the
smallest discrepancies between the corresponding universal factors $\hat c$. 

The effect of varying the perturbative matching point in the optimal schemes
is described by Table~\ref{tab:phiratio_res}. We see that numbers
are very stable for $n\ge 6$, while the uncertainty increases
with $n$ due to progressive error accumulation at each step. Final results, reported in
the second column of Table~\ref{tab:fitcoefs} refer to $n=7$. Note that typical 
relative errors are as big as~5\%, which may result in a sizeable error in hadronic
matrix elements, solely due to renormalisation.

\section{Connection to hadronic observables}

Having computed the universal evolution factors $\hat c(\mu_{\rm had})$, which 
provide the RG-running from the low energy matching scale $\mu_{\rm had}$
to a formally infinite one, we proceed to establish the connection between bare 
lattice operators and their RGI counterparts. The latter, defined in Eq.~(\ref{eq:rgi})
from the integration of the Callan-Symanzik equation, are related to the
bare operators used in lattice simulations via a total renormalisation factor 
$Z_{\rm\scriptscriptstyle RGI}(g_0)$, defined as
\begin{gather}
\label{eq:ZRGI}
Z_{\rm\scriptscriptstyle RGI}(g_0) = Z(g_0,a\mu_{\rm had})\hat c(\mu_{\rm had}) \ .
\end{gather}
The $Z_{\rm\scriptscriptstyle RGI}$ factor does not depend on any renormalisation 
scale and carries a dependence upon the renormalisation condition only via cutoff 
effects. 

\begin{table}[!t]
    \footnotesize
    \centering
    \renewcommand{\arraystretch}{1}
    \def\onelp{\small{1-lp}}
    \begin{tabular}{ccclcccc} \Hline\\[-1.0ex]
      $\beta$  & $\kappa$  & $L/a$ & $ \gbar^2_{\rm\scriptscriptstyle SF}(L)$ & $Z^{+}$ & $Z^{-}$ \\[1.5ex] \hline\\[-2.0ex]
      5.20 & 0.13600 & 4 & 3.65(3)    & 0.7547(19) & 0.4797(12) \\
           &         & 6 & 4.61(4)    & 0.7715(20) & 0.4383(11)  \\[0.7ex] \hline \\[-1.0ex]
      5.29 & 0.13641 & 4 & 3.394(17)  & 0.7558(17) & 0.5070(11) \\
           &         & 6 & 4.297(37)  & 0.7749(24) & 0.4644(13) \\
           &         & 8 & 5.65(9)    & 0.8036(26) & 0.4339(12) \\[0.7ex] \hline \\[-1.0ex]
      5.40 & 0.13669 & 4 & 3.188(24)  & 0.7591(16) & 0.5342(11) \\
           &         & 6 & 3.864(34)  & 0.7709(21) & 0.4871(13) \\
           &         & 8 & 4.747(63)  & 0.7938(22) & 0.4583(11)  \\[1.0ex]
           \Hline
    \end{tabular}
    \caption{\small 
      Results for $Z^+$ and $Z^-$ with $c_{\rm t}$ set to its 2--loop value.
      The values of $\gbar^2_{\rm\scriptscriptstyle SF}$ are from \cite{alpha:Nf2_2}.
      The hopping parameters $\kappa$ used in the simulations are the
      critical ones ($\kappa_{\rm cr}$) of \cite{Gockeler:2004rp}.
    }\label{tab:renconsts1}
\vskip 0.8cm
    \footnotesize
    \centering
    \renewcommand{\arraystretch}{1}
    \begin{tabular}{ccclcccc} \Hline\\[-1.0ex]
      $\beta$  & $\kappa$  & $L/a$ & $ \gbar^2_{\rm\scriptscriptstyle SF}(L)$ & ${\cal Z}^{(1)}$ & ${\cal Z}^{(2)}$ & ${\cal Z}^{(3)}$ & ${\cal Z}^{(4)}$ \\[1.5ex] \hline\\[-2.0ex]
      5.20 & 0.13600 & 4 & 3.65(3)    & 0.7793(17) & 0.9741(16) & 0.8681(16) & 0.8317(13) \\
           &         & 6 & 4.61(4)    & 0.7118(17) & 0.9409(15) & 0.7857(13) & 0.7921(13) \\[0.7ex] \hline \\[-1.0ex]
      5.29 & 0.13641 & 4 & 3.394(17)  & 0.7862(16) & 0.9766(16) & 0.8768(15) & 0.8397(14) \\
           &         & 6 & 4.297(37)  & 0.7275(20) & 0.9431(18) & 0.7992(16) & 0.8017(15) \\
           &         & 8 & 5.65(9)    & 0.6612(19) & 0.9150(16) & 0.7337(15) & 0.7619(14) \\[0.7ex] \hline \\[-1.0ex]
      5.40 & 0.13669 & 4 & 3.188(24)  & 0.7972(15) & 0.9805(14) & 0.8864(14) & 0.8497(12) \\
           &         & 6 & 3.864(34)  & 0.7378(18) & 0.9434(17) & 0.8098(15) & 0.8094(14) \\
           &         & 8 & 4.747(63)  & 0.6840(16) & 0.9231(14) & 0.7529(13) & 0.7781(12) \\[1.0ex]
           \Hline
    \end{tabular}
    \caption{\small
      Results for ${\cal Z}^{(k)}$ with $c_{\rm t}$ set to its 2--loop value.
      The values of $\gbar^2_{\rm\scriptscriptstyle SF}$ are from \cite{alpha:Nf2_2}.
      The hopping parameters $\kappa$ used in the simulations are the
      critical ones ($\kappa_{\rm cr}$) of \cite{Gockeler:2004rp}.
    }\label{tab:renconsts2}
\end{table}

In order to obtain $Z(g_0,a\mu_{\rm had})$, we
follow~\cite{DellaMorte:2005kg} and compute $Z(g_0,a\mu)$ at three 
values of the lattice spacing, namely $\beta=\{5.20,5.29,5.40\}$, which belong
to a range of inverse couplings commonly used for simulations of two-flavour QCD 
in physically large volumes. Simulation parameters and results are collected in 
Table~\ref{tab:renconsts1} for the relativistic operators $Q_1^\pm$ and in 
Table~\ref{tab:renconsts2} for the static ones ${\cal Q'}_k^{+}$. 

While the simulation at $(\beta=5.20,L/a=6)$ is exactly at the
target value for $\gbar_{\rm\scriptscriptstyle SF}^2(L_{\rm max})$, 
corresponding to $Z(g_0,a\mu_{\rm had})$, the simulations at the other $\beta$
values require a slight interpolation. We adopt a 
fit ansatz, motivated by Eq.(\ref{eq:rgi}),
\begin{equation}
\ln\left(Z\right)=c_1+c_2\ln(\gbar_{\rm\scriptscriptstyle SF}^2) \;,
\label{logfit_zastat}
\end{equation}
in order to interpolate the $Z$ factors between the values of 
$\gbar^2_{\rm\scriptscriptstyle SF}$ straddling the target value 
$\gbar_{\rm\scriptscriptstyle SF}^2(L_{\rm max}) = 4.61$. 
Note that the fits take into account the (independent) errors of both $Z$ and 
$\gbar_{\rm\scriptscriptstyle SF}^2$. Moreover, we have conservatively 
augmented the fit errors by the difference between the fit results
of Eq.~(\ref{logfit_zastat}) and the results from a naive two-point linear
interpolation in $\gbar_{\rm\scriptscriptstyle SF}^2$.
The coefficients $c_2$ of the fits (\ref{logfit_zastat})
deviate in a range of $7\%-30\%$ from the lowest order coefficients
$\gamma_0/(2b_0)$, signalling the presence of moderate 
higher-order perturbative effects. 

\begin{table}[!t]
  \footnotesize
  \centering
  \renewcommand{\arraystretch}{1}
  \begin{tabular}{cccccc} \Hline\\[-1.5ex]
    $\beta$ & $Z^{+}$ & $Z^{+}_{\rm\scriptscriptstyle RGI}$ & & $Z^{-}$ & $Z^{-}_{\rm\scriptscriptstyle RGI}$ \\[1.0ex] \hline\\[-2.5ex]
     5.20  & 0.7715(20)  & 0.927(2)(55)  & & 0.4383(11) &  0.241(1)(10)  \\
     5.29  & 0.7825(27)  & 0.940(3)(56)  & & 0.4560(23) &  0.251(1)(10)  \\
     5.40  & 0.7905(26)  & 0.950(3)(56)  & & 0.4623(25) &  0.255(1)(10)  \\[1.0ex] \hline\\[-1.7ex]
    $\beta$ & ${\cal Z}^{(1)}$ & ${\cal Z}^{(1)}_{\rm\scriptscriptstyle RGI}$ & & ${\cal Z}^{(2)}$  & ${\cal Z}^{(2)}_{\rm\scriptscriptstyle RGI}$ \\[1.0ex] \hline\\[-2.5ex]
     5.20  & 0.7118(17)  & 0.548(1)(28)  & & 0.9409(15)  & 0.718(1)(26)  \\
     5.29  & 0.7093(27)  & 0.546(2)(28)  & & 0.9374(30)  & 0.715(2)(26)  \\
     5.40  & 0.6904(40)  & 0.532(3)(28)  & & 0.9233(46)  & 0.704(4)(26)  \\[1.0ex] \hline\\[-1.5ex]
    $\beta$ & ${\cal Z}^{(3)}$ &  ${\cal Z}^{(3)}_{\rm\scriptscriptstyle RGI}$ & & ${\cal Z}^{(4)}$ & ${\cal Z}^{(4)}_{\rm\scriptscriptstyle RGI}$ \\[1.0ex] \hline\\[-2.5ex]
     5.20  & 0.7857(13)  & 0.423(1)(15)  & & 0.7921(13)  & 0.631(1)(17)  \\
     5.29  & 0.7836(40)  & 0.422(2)(15)  & & 0.7916(18)  & 0.630(1)(17)  \\
     5.40  & 0.7567(75)  & 0.408(4)(14)  & & 0.7807(38)  & 0.621(3)(17)  \\[1.0ex] \Hline\\[-1.0ex]
  \end{tabular}
  \caption{\small 
    Results for $Z^{+}$, $Z^{-}$, ${\cal Z}^{(k)}$ and $Z^{+}_{\rm\scriptscriptstyle RGI}$, $Z^{-}_{\rm\scriptscriptstyle RGI}$, ${\cal Z}^{(k)}_{\rm\scriptscriptstyle RGI}$ for three bare gauge coupling values
    corresponding to our low-energy matching point at $\bar g^2_{\rm\scriptscriptstyle SF}=4.61$ in the SF
    scheme.
  }\label{tab:ZRGIfinalnumbers}
\end{table}

The resulting numbers for the renormalisation factors at the low energy matching
scale, and also for the RGI renormalisation factors $Z_{\rm\scriptscriptstyle RGI}(g_0)$, 
are collected in Table~\ref{tab:ZRGIfinalnumbers}. The first error of the
$Z_{\rm\scriptscriptstyle RGI}$'s stems from the error of $Z$ factors, whereas 
the second accounts for the uncertainties in the universal factors $\hat c$. 
Note that only the first of these errors should be added in quadrature to the error
of the bare hadronic matrix elements, once these become available from future computations, 
in order to obtain the total error of the renormalised quantity, at a given lattice spacing. 
The second error, which is entirely unrelated to the discretisation of the theory, should only be added
in quadrature to the continuum extrapolated hadronic matrix element.
For the sake of convenience, a representation of the numerical results of 
Table~\ref{tab:ZRGIfinalnumbers} by interpolating polynomials is also adopted, i.e. 
\begin{equation}
\label{eq:fitpol}
Z_{\rm\scriptscriptstyle RGI} = a_0 + a_1(\beta - 5.2) + a_2(\beta-5.2)^2\ , 
\end{equation}
which can be used at any intermediate value of $\beta$ between $\beta=5.20$ and
$\beta=5.40$. Fit coefficients are reported in Table~\ref{tab:fitcoefs} for the 
various operators. The uncertainty of the RGI constants at intermediate points may 
be easily obtained from those at the simulation points, see Table~\ref{tab:ZRGIfinalnumbers}, 
by linear interpolation. 

As a final remark, we observe that the simulation of the renormalisation factors
at $\beta=5.20,L/a=4$ is not at the target value for $\bar g^2_{\rm\scriptscriptstyle SF}$. 
We used is as a check of the independence of the $Z_{\rm\scriptscriptstyle RGI}$,
computed via Eq.~(\ref{eq:ZRGI}) from the low energy matching scale. Specifically, 
the two measured values of $Z$-factors at $\beta=5.20$ have been used in order to 
extrapolate the renormalisation constants at  
$\bar g_{\rm\scriptscriptstyle SF}^2(L_{\rm max}/2) = 3.0318$, where the non-perturbative
matching with the universal evolution factors $\hat c$ has been subsequently performed. 
Results turned out to be fully compatible with those quoted in Table~\ref{tab:ZRGIfinalnumbers}.

\section{Conclusions}

Using standard SF methods, we have performed a fully non-perturbative
computation of the renormalisation and RG running of several four-fermion 
operators in $\NF=2$ QCD. We have considered the two operators made of 
four relativistic quark fields with a left-left Dirac structure and the 
complete basis of operators with two static and two relativistic quarks.  
The Wilson lattice actions have been implemented for both the gauge and the 
fermionic parts, the latter with a non-perturbatively tuned Clover term. 
The HYP2 discretisation of the static quark turned out to be the less noisy 
choice, after comparison with other options. Only the parity-odd parts of 
the operators have been analysed, as their renormalisation pattern is 
unaffected due to the loss of chiral symmetry by the regularisation. 

\begin{table}[!t]
    \footnotesize
    \centering
    \renewcommand{\arraystretch}{1}
    \begin{tabular}{ccccc} \Hline\\[-1.5ex]
      $Q$                & $\hat c(L_{\rm max}^{-1})$ & $a_0$  & $a_1$ & $a_2$ \\[1.0ex] \hline\\[-2.0ex]
      $Q_1^+$            & 1.202(71)                  & 0.9270 & 0.1741  & -0.2973  \\[0.7ex]
      $Q_1^-$            & 0.551(22)                  & 0.2414 & 0.1431  & -0.3853  \\[0.7ex]
      ${\cal Q'}_1^{+}$  & 0.770(35)                  & 0.5481 & 0.0285  & -0.5546  \\[0.7ex]
      ${\cal Q'}_2^{+}$  & 0.763(24)                  & 0.7179 & 0.0010  & -0.3407  \\[0.7ex]
      ${\cal Q'}_3^{+}$  & 0.539(18)                  & 0.4235 & 0.0411  & -0.5962  \\[0.7ex]
      ${\cal Q'}_4^{+}$  & 0.796(20)                  & 0.6305 & 0.0291  & -0.3723  \\[1.0ex]
      \Hline
    \end{tabular}
    \caption{\small
      Universal factors $\hat c$ and coefficients of the interpolating polynomials of the 
      RGI renormalisation constants, see Eq.~(\ref{eq:fitpol}). Uncertainties are discussed in the text. 
    }\label{tab:fitcoefs}
\end{table}

Our results are an essential building block for
any $\NF=2$ computation of quantities like $B_K$ and $B_B$. Nevertheless,
their precision is somewhat limited by increased statistical fluctuations at
the three strongest couplings and by the lack of a fourth, finer, lattice
resolution which would
improve the continuum extrapolation of the operator SSFs. This could lead to a
potentially unsatisfactory total error on hadronic matrix elements. Future refinement (besides
using a two--loop estimate of $c_{\rm t}$ throughout the runs and increased
statistics at the three strongest couplings) is necessary, either
by simulating closer to the continuum limit, or by completely
removing leading order discretisation effects from the simulations.

\section*{Acknowledgements}
We wish to thank M.~Della~Morte, R.~Frezzotti, F.~Knechtli, M.~L\"uscher,
S.~Sint and R.~Sommer for help and useful discussions.
We thank the authors of ref.~\cite{DellaMorte:2005kg} for making their 
$N_{\rm f}=2$ dynamical quark configuration ensemble available to us. 
G.H.~acknowledges partial financial support from the DFG. F.P. acknowledges 
the Alexander-von-Humboldt Foundation for partial financial support. 
M.P. acknowledges financial support by an EIF Marie Curie fellowship of the 
European Community's Sixth Framework Programme under contract number 
MEIF-CT-2006-040458. C.P. acknowledges financial support by the Ram\'on y 
Cajal programme and by the CICyT project FPA2006-05807.
This work was supported in part by the EU Contract  No. MRTN-CT-2006-035482, 
``FLAVIAnet''. We are grateful to the APE team at DESY-Zeuthen for their 
constant support and for use of the APE computers, on which our simulations 
were performed.

\begin{table}[!ht]{
    \footnotesize
    \centering
    \begin{tabular}{lrlrccc}
      \Hline \\[-1.2ex]
      $\gbar^2_{\rm\scriptscriptstyle SF}(L)$ & $~~~~~\beta~~~~$ & $~~~~~\hopc$ & ${L}/{a}$ & ${Z}^{+}\left(g_0,{a}/{L}\right)$ & ${Z}^{+}\left(g_0,{a}/{2L}\right)$ &\
      ${\Sigma}^{+}\left(g_0,{a}/{L}\right)$ \\[1.0ex]
      \hline \\[-2.0ex]
      0.9793 &  9.50000  & 0.131532  &  6   & 0.8714(14) & 0.8827(22) & 1.0129(30)  \\
             &  9.73410  & 0.131305  &  8   & 0.8765(16) & 0.8852(25) & 1.0099(34)  \\
             &  10.05755 & 0.131069  &  12  & 0.8899(17) & 0.9022(52) & 1.0138(61)  \\
      \hline \\[-2.0ex]                        
      1.1814 &  8.50000  & 0.132509  &  6   & 0.8510(14) & 0.8683(48) & 1.0204(58)  \\
             &  8.72230  & 0.132291  &  8   & 0.8594(29) & 0.8849(33) & 1.0296(52)  \\
             &  8.99366  & 0.131975  &  12  & 0.8753(20) & 0.9019(64) & 1.0304(77)  \\
      \hline \\[-2.0ex]                        
      1.5078 &  7.54200  & 0.133705  &  6   & 0.8309(18) & 0.8580(47) & 1.0327(60)  \\
             &  7.72060  & 0.133497  &  8   & 0.8395(38) & 0.8725(62) & 1.0392(87)  \\
      1.5031 &  7.50000  & 0.133815  &  6   & 0.8317(15) & 0.8390(55) & 1.0088(69)  \\
             &  8.02599  & 0.133063  &  12  & 0.8531(44) & 0.8811(83) & 1.0328(111) \\
      \hline \\[-2.0ex]                        
      2.0142 &  6.60850  & 0.135260  &  6   & 0.8023(19) & 0.8382(32) & 1.0448(47)  \\
             &  6.82170  & 0.134891  &  8   & 0.8209(40) & 0.8545(45) & 1.0410(74)  \\
             &  7.09300  & 0.134432  &  12  & 0.8400(44) & 0.8771(70) & 1.0442(100) \\
      \hline \\[-2.0ex]                        
      2.4792 &  6.13300  & 0.136110  &  6   & 0.7885(33) & 0.8371(71)  & 1.0616(100) \\
             &  6.32290  & 0.135767  &  8   & 0.8038(31) & 0.8466(127) & 1.0531(163) \\
             &  6.63164  & 0.135227  &  12  & 0.8290(39) & 0.8921(148) & 1.0761(186) \\
      \hline \\[-2.0ex]                        
      3.3340 &  5.62150  & 0.136665  &  6   & 0.7667(45) & 0.8193(129) & 1.0686(179) \\
             &  5.80970  & 0.136608  &  8   & 0.7927(45) & 0.8812(126) & 1.1116(171) \\
             &  6.11816  & 0.136139  &  12  & 0.8252(68) & 0.9040(110) & 1.0955(161) \\[1.0ex]
      \hline\hline\\[-1.2ex]
      $\gbar^2_{\rm\scriptscriptstyle SF}(L)$ & $~~~~~\beta~~~~$ & $~~~~~\hopc$ & ${L}/{a}$ & ${Z}^{-}\left(g_0,{a}/{L}\right)$ & $\mathcal{Z}^{-}\left(g_0,{a}/{2L}\right)$ &\
      ${\Sigma}^{-}\left(g_0,{a}/{L}\right)$ \\[1.0ex]
      \hline\\[-2.0ex]
      0.9793 &  9.50000  & 0.131532  &  6   & 0.7841(12) & 0.7546(15) & 0.9623(24)  \\
             &  9.73410  & 0.131305  &  8   & 0.7767(10) & 0.7500(27) & 0.9657(37)  \\
             &  10.05755 & 0.131069  &  12  & 0.7696(09)  & 0.7491(27) & 0.9733(36)  \\
      \hline \\[-2.0ex] 
      1.1814 &  8.50000  & 0.132509  &  6   & 0.7512(11) & 0.7185(34) & 0.9564(47)  \\
             &  8.72230  & 0.132291  &  8   & 0.7461(18) & 0.7180(19) & 0.9623(35)  \\
             &  8.99366  & 0.131975  &  12  & 0.7372(10) & 0.7075(30) & 0.9598(43)  \\
      \hline \\[-2.0ex] 
      1.5078 &  7.54200  & 0.133705  &  6   & 0.7091(11) & 0.6696(24) & 0.9443(37)  \\
             &  7.72060  & 0.133497  &  8   & 0.6998(18) & 0.6680(44) & 0.9547(68)  \\
      1.5031 &  7.50000  & 0.133815  &  6   & 0.7062(09) & 0.6655(25) & 0.9424(38)  \\
             &  8.02599  & 0.133063  &  12  & 0.6954(25) & 0.6584(31) & 0.9468(56)  \\
      \hline \\[-2.0ex] 
      2.0142 &  6.60850  & 0.135260  &  6   & 0.6475(13) & 0.5965(16) & 0.9212(31)  \\
             &  6.82170  & 0.134891  &  8   & 0.6428(27) & 0.6011(25) & 0.9351(55)  \\
             &  7.09300  & 0.134432  &  12  & 0.6379(22) & 0.5898(29) & 0.9246(55)  \\
      \hline \\[-2.0ex] 
      2.4792 &  6.13300  & 0.136110  &  6   & 0.6029(21) & 0.5470(37) & 0.9072(69)  \\
             &  6.32290  & 0.135767  &  8   & 0.5994(16) & 0.5336(38) & 0.8902(68)  \\
             &  6.63164  & 0.135227  &  12  & 0.5995(22) & 0.5386(53) & 0.8984(94)  \\
      \hline \\[-2.0ex] 
      3.3340 &  5.62150  & 0.136665  &  6   & 0.5288(31) & 0.4610(69) & 0.8718(140) \\
             &  5.80970  & 0.136608  &  8   & 0.5363(24) & 0.4632(63) & 0.8637(124) \\
             &  6.11816  & 0.136139  &  12  & 0.5417(31) & 0.4698(49) & 0.8672(102) \\[0.5ex]
             \Hline \\[-1.5ex]                         
    \end{tabular}
    \vskip -0.2cm
    \caption{\small 
      Numerical values of the renormalisation constants $Z^+$, $Z^-$
      and the step scaling functions ${\Sigma}^+$, $\Sigma^-$ at various
      renormalised SF couplings and lattice spacings. Data at 
      $\gbar^2_{\rm\scriptscriptstyle SF}=0.9793,\ 1.1814,\ 1.5031$ have been obtained 
      with $c_{\rm t}$ evaluated in one-loop perturbation theory. The remaining data 
      have been obtained with $c_{\rm t}$ evaluated in two-loop perturbation theory. 
    }
    \label{tab:tab2}
  }
\end{table}

\begin{table}[!t]{
    \footnotesize
    \centering
    \begin{tabular}{lrlrccc}
      \Hline \\[-1.2ex]
      $\gbar^2_{\rm\scriptscriptstyle SF}(L)$ & $~~~~~\beta~~~~$ & $~~~~~\hopc$ & ${L}/{a}$ & $\mathcal{Z}^{(1)}\left(g_0,{a}/{L}\right)$ & $\mathcal{Z}^{(1)}\left(g_0,{a}/{2L}\right)$ &\
      ${\Sigma}^{(1)}\left(g_0,{a}/{L}\right)$ \\[1.0ex]
      \hline \\[-2.0ex]
      0.9793 &  9.50000  & 0.131532  &  6   &  0.8958(15) & 0.8630(18) & 0.9634(25) \\
             &  9.73410  & 0.131305  &  8   &  0.8845(13) & 0.8486(23) & 0.9594(29) \\
             &  10.05755 & 0.131069  &  12  &  0.8733(15) & 0.8335(38) & 0.9545(47) \\
      \hline \\[-2.0ex]                         
      1.1814 &  8.50000  & 0.132509  &  6   &  0.8771(16) & 0.8421(41) & 0.9601(50) \\
             &  8.72230  & 0.132291  &  8   &  0.8650(22) & 0.8304(29) & 0.9600(41) \\
             &  8.99366  & 0.131975  &  12  &  0.8503(17) & 0.8117(47) & 0.9545(59) \\
      \hline \\[-2.0ex]                         
      1.5078 &  7.54200  & 0.133705  &  6   &  0.8531(16) & 0.8043(29) & 0.9428(39) \\
             &  7.72060  & 0.133497  &  8   &  0.8385(34) & 0.7924(53) & 0.9450(74) \\
      1.5031 &  7.50000  & 0.133815  &  6   &  0.8547(13) & 0.8161(43) & 0.9548(52) \\
             &  8.02599  & 0.133063  &  12  &  0.8161(43) & 0.7638(37) & 0.9359(67) \\
      \hline \\[-2.0ex]                         
      2.0142 &  6.60850  & 0.135260  &  6   &  0.8190(17) & 0.7535(22) & 0.9200(33) \\
             &  6.82170  & 0.134891  &  8   &  0.8082(31) & 0.7334(35) & 0.9075(56) \\
             &  7.09300  & 0.134432  &  12  &  0.7798(28) & 0.7102(38) & 0.9108(59) \\
      \hline \\[-2.0ex]                         
      2.4792 &  6.13300  & 0.136110  &  6   &  0.7937(27) & 0.7085(49) & 0.8927(68) \\
             &  6.32290  & 0.135767  &  8   &  0.7754(21) & 0.6841(90) & 0.8823(119) \\
             &  6.63164  & 0.135227  &  12  &  0.7492(25) & 0.6691(65) & 0.8931(92) \\
      \hline \\[-2.0ex]                         
      3.3340 &  5.62150  & 0.136665  &  6   &  0.7570(38) & 0.6233(67) & 0.8235(98) \\
             &  5.80970  & 0.136608  &  8   &  0.7330(37) & 0.5987(71) & 0.8168(106) \\
             &  6.11816  & 0.136139  &  12  &  0.7048(46) & 0.5658(65) & 0.8028(106) \\[1.0ex]
      \hline\hline\\[-1.2ex]
      $\gbar^2_{\rm\scriptscriptstyle SF}(L)$ & $~~~~~\beta~~~~$ & $~~~~~\hopc$ & ${L}/{a}$ & $\mathcal{Z}^{(2)}\left(g_0,{a}/{L}\right)$ & $\mathcal{Z}^{(2)}\left(g_0,{a}/{2L}\right)$ &\
      ${\Sigma}^{(2)}\left(g_0,{a}/{L}\right)$ \\[1.0ex]
      \hline\\[-2.0ex]
      0.9793 &  9.50000  & 0.131532  &  6   & 0.9810(11) &  0.9645(15) & 0.9832(19)  \\
             &  9.73410  & 0.131305  &  8   & 0.9739(10) &  0.9613(23) & 0.9871(26)  \\
             &  10.05755 & 0.131069  &  12  & 0.9676(09) &  0.9627(37) & 0.9949(39)  \\
      \hline \\[-2.0ex] 
      1.1814 &  8.50000  & 0.132509  &  6   & 0.9769(13) &  0.9599(42) & 0.9826(45)  \\
             &  8.72230  & 0.132291  &  8   & 0.9699(18) &  0.9572(21) & 0.9869(28)  \\
             &  8.99366  & 0.131975  &  12  & 0.9644(11) &  0.9516(29) & 0.9867(32)  \\
      \hline \\[-2.0ex] 
      1.5078 &  7.54200  & 0.133705  &  6   & 0.9738(11) &  0.9507(25) & 0.9762(28)  \\
             &  7.72060  & 0.133497  &  8   & 0.9662(24) &  0.9482(40) & 0.9813(48)  \\
      1.5031 &  7.50000  & 0.133815  &  6   & 0.9707(11) &  0.9579(24) & 0.9868(28)  \\
             &  8.02599  & 0.133063  &  12  & 0.9579(24) &  0.9392(46) & 0.9805(54)  \\
      \hline \\[-2.0ex] 
      2.0142 &  6.60850  & 0.135260  &  6   & 0.9667(12) &  0.9371(18) & 0.9694(22)  \\
             &  6.82170  & 0.134891  &  8   & 0.9551(24) &  0.9372(23) & 0.9813(34)  \\
             &  7.09300  & 0.134432  &  12  & 0.9476(22) &  0.9179(49) & 0.9687(56)  \\
      \hline \\[-2.0ex] 
      2.4792 &  6.13300  & 0.136110  &  6   & 0.9605(23) &  0.9207(48) & 0.9586(55)  \\
             &  6.32290  & 0.135767  &  8   & 0.9496(14) &  0.9099(51) & 0.9582(56)  \\
             &  6.63164  & 0.135227  &  12  & 0.9385(23) &  0.8995(56) & 0.9584(64)  \\
      \hline \\[-2.0ex] 
      3.3340 &  5.62150  & 0.136665  &  6   & 0.9511(41) &  0.8997(75) & 0.9459(89)  \\
             &  5.80970  & 0.136608  &  8   & 0.9352(37) &  0.8769(84) & 0.9376(97)  \\
             &  6.11816  & 0.136139  &  12  & 0.9200(35) &  0.8669(59) & 0.9423(74)  \\[0.5ex]
             \Hline \\[-1.5ex]                         
    \end{tabular}
    \vskip -0.2cm
    \caption{\small
      Numerical values of the renormalisation constants $\mathcal{Z}^{(1)}$, $\mathcal{Z}^{(2)}$ 
      and the step-scaling functions ${\Sigma}^{(1)}$, ${\Sigma}^{(2)}$ with HYP2 action at various
      renormalised SF couplings and lattice spacings. Data at 
      $\gbar^2_{\rm\scriptscriptstyle SF}=0.9793,\ 1.1814,\ 1.5031$ have been obtained 
      with $c_{\rm t}$ evaluated in one-loop perturbation theory. The remaining data have been 
      obtained with $c_{\rm t}$ evaluated in two-loop perturbation theory. 
    }
    \label{tab:tab3}
  }
\end{table}

\begin{table}[!t]{
    \footnotesize
    \centering
    \begin{tabular}{lrlrccc}
      \Hline \\[-1.2ex]
      $\gbar^2_{\rm\scriptscriptstyle SF}(L)$ & $~~~~~\beta~~~~$ & $~~~~~\hopc$ & ${L}/{a}$ & $\mathcal{Z}^{(3)}\left(g_0,{a}/{L}\right)$ & $\mathcal{Z}^{(3)}\left(g_0,{a}/{2L}\right)$ &\
      ${\Sigma}^{(3)}\left(g_0,{a}/{L}\right)$ \\[1.0ex]
      \hline \\[-2.0ex]
      0.9793 &  9.50000  & 0.131532  &  6   & 0.9306(10) &  0.8879(13) &  0.9541(17)   \\
             &  9.73410  & 0.131305  &  8   & 0.9164(08) &  0.8762(12) &  0.9561(16)  \\
             &  10.05755 & 0.131069  &  12  & 0.8973(07) &  0.8586(21) &  0.9569(24)  \\
      \hline \\[-2.0ex]
      1.1814 &  8.50000  & 0.132509  &  6   & 0.9179(10) &  0.8715(30) &  0.9494(34)  \\
             &  8.72230  & 0.132291  &  8   & 0.8996(13) &  0.8526(17) &  0.9478(23)  \\
             &  8.99366  & 0.131975  &  12  & 0.8810(09) &  0.8333(27) &  0.9458(33)  \\
      \hline \\[-2.0ex]
      1.5078 &  7.54200  & 0.133705  &  6   & 0.9038(10) &  0.8411(20) &  0.9307(25)  \\
             &  7.72060  & 0.133497  &  8   & 0.8814(20) &  0.8207(26) &  0.9311(36)  \\
      1.5031 &  7.50000  & 0.133815  &  6   & 0.8998(10) &  0.8564(20) &  0.9517(25)  \\
             &  8.02599  & 0.133063  &  12  & 0.8564(20) &  0.7976(36) &  0.9314(48)  \\
      \hline \\[-2.0ex]
      2.0142 &  6.60850  & 0.135260  &  6   & 0.8794(12) &  0.7981(16) &  0.9075(22)  \\
             &  6.82170  & 0.134891  &  8   & 0.8543(22) &  0.7788(22) &  0.9116(34)  \\
             &  7.09300  & 0.134432  &  12  & 0.8231(22) &  0.7405(37) &  0.8997(50)  \\
      \hline \\[-2.0ex]
      2.4792 &  6.13300  & 0.136110  &  6   & 0.8596(21) &  0.7582(42) &  0.8821(53)  \\
             &  6.32290  & 0.135767  &  8   & 0.8347(13) &  0.7329(52) &  0.8780(63)  \\
             &  6.63164  & 0.135227  &  12  & 0.7972(19) &  0.6999(50) &  0.8780(66)  \\
      \hline \\[-2.0ex]
      3.3340 &  5.62150  & 0.136665  &  6   & 0.8346(36) &  0.7043(62) &  0.8439(83)  \\
             &  5.80970  & 0.136608  &  8   & 0.7996(33) &  0.6568(63) &  0.8214(85)  \\
             &  6.11816  & 0.136139  &  12  & 0.7587(30) &  0.6241(58) &  0.8227(82)  \\[1.0ex]
      \hline\hline\\[-1.2ex]
      $\gbar^2_{\rm\scriptscriptstyle SF}(L)$ & $~~~~~\beta~~~~$ & $~~~~~\hopc$ & ${L}/{a}$ & $\mathcal{Z}^{(4)}\left(g_0,{a}/{L}\right)$ & $\mathcal{Z}^{(4)}\left(g_0,{a}/{2L}\right)$ &\
      ${\Sigma}^{(4)}\left(g_0,{a}/{L}\right)$ \\[1.0ex]
      \hline\\[-2.0ex]
      0.9793 &  9.50000  & 0.131532  &  6   & 0.9281(09) &  0.9082(12) &  0.9786(16)  \\
             &  9.73410  & 0.131305  &  8   & 0.9213(08) &  0.9033(12) &  0.9804(15)  \\
             &  10.05755 & 0.131069  &  12  & 0.9147(06) &  0.9009(21) &  0.9849(23)  \\
      \hline \\[-2.0ex]
      1.1814 &  8.50000  & 0.132509  &  6   & 0.9158(10) &  0.8950(24) &  0.9773(28)  \\
             &  8.72230  & 0.132291  &  8   & 0.9094(13) &  0.8904(14) &  0.9791(21)  \\
             &  8.99366  & 0.131975  &  12  & 0.9022(08) &  0.8811(19) &  0.9766(23)  \\
      \hline \\[-2.0ex]
      1.5078 &  7.54200  & 0.133705  &  6   & 0.8995(10) &  0.8691(18) &  0.9662(23)  \\
             &  7.72060  & 0.133497  &  8   & 0.8924(18) &  0.8668(28) &  0.9714(37)  \\
      1.5031 &  7.50000  & 0.133815  &  6   & 0.8979(10) &  0.8829(16) &  0.9833(2)   \\
             &  8.02599  & 0.133063  &  12  & 0.8829(16) &  0.8536(28) &  0.9668(36)  \\
      \hline \\[-2.0ex]
      2.0142 &  6.60850  & 0.135260  &  6   & 0.8749(11) &  0.8353(14) &  0.9548(20)  \\
             &  6.82170  & 0.134891  &  8   & 0.8671(19) &  0.8321(18) &  0.9597(30)  \\
             &  7.09300  & 0.134432  &  12  & 0.8562(18) &  0.8143(31) &  0.9510(42)  \\
      \hline \\[-2.0ex]
      2.4792 &  6.13300  & 0.136110  &  6   & 0.8566(19) &  0.8006(36) &  0.9347(47)  \\
             &  6.32290  & 0.135767  &  8   & 0.8461(14) &  0.7906(41) &  0.9344(51)  \\
             &  6.63164  & 0.135227  &  12  & 0.8360(17) &  0.7832(46) &  0.9368(58)  \\
      \hline \\[-2.0ex]
      3.3340 &  5.62150  & 0.136665  &  6   & 0.8268(35) &  0.7474(58) &  0.9040(80)  \\
             &  5.80970  & 0.136608  &  8   & 0.8142(30) &  0.7337(63) &  0.9011(85)  \\
             &  6.11816  & 0.136139  &  12  & 0.8002(26) &  0.7239(48) &  0.9046(67)  \\[0.5ex]
             \Hline \\[-1.5ex]                         
    \end{tabular}
    \vskip -0.2cm
    \caption{\small 
      Numerical values of the renormalisation constants $\mathcal{Z}^{(3)}$, $\mathcal{Z}^{(4)}$ 
      and the step-scaling functions ${\Sigma}^{(3)}$, ${\Sigma}^{(4)}$ with HYP2 action at various
      renormalised SF couplings and lattice spacings. Data at 
      $\gbar^2_{\rm\scriptscriptstyle SF}=0.9793,\ 1.1814,\ 1.5031$ have been obtained 
      with $c_{\rm t}$ evaluated in one-loop perturbation theory. The remaining data have been 
      obtained with $c_{\rm t}$ approximated in two-loop perturbation theory. 
    }
    \label{tab:tab4}
  }
\end{table}

\newpage

\newpage

\newpage 


\newpage

\begin{figure}[!hb]
  \begin{center}
    \vskip 1.5cm
     \hskip -0.2cm \epsfig{figure=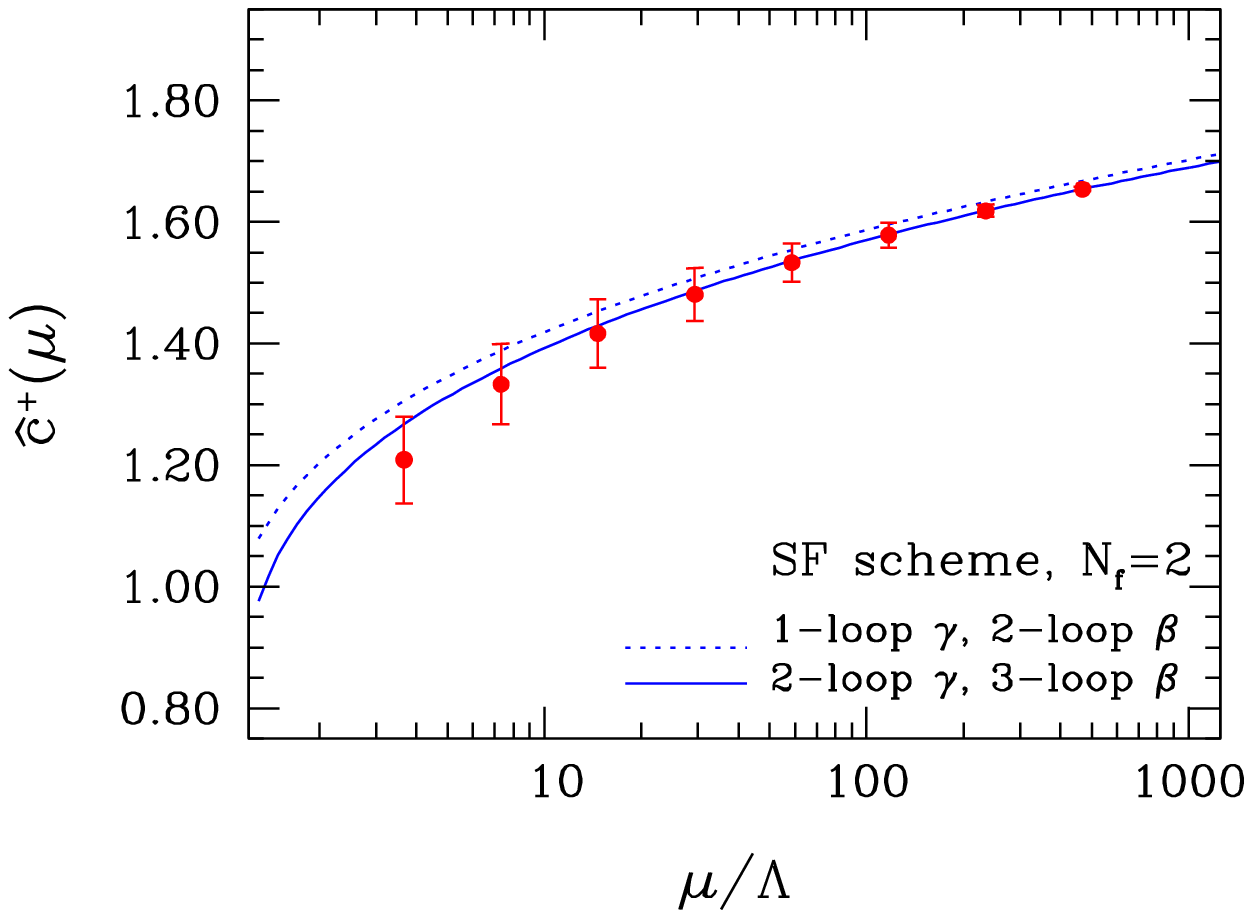, width=6.5 true cm}
     \epsfig{figure=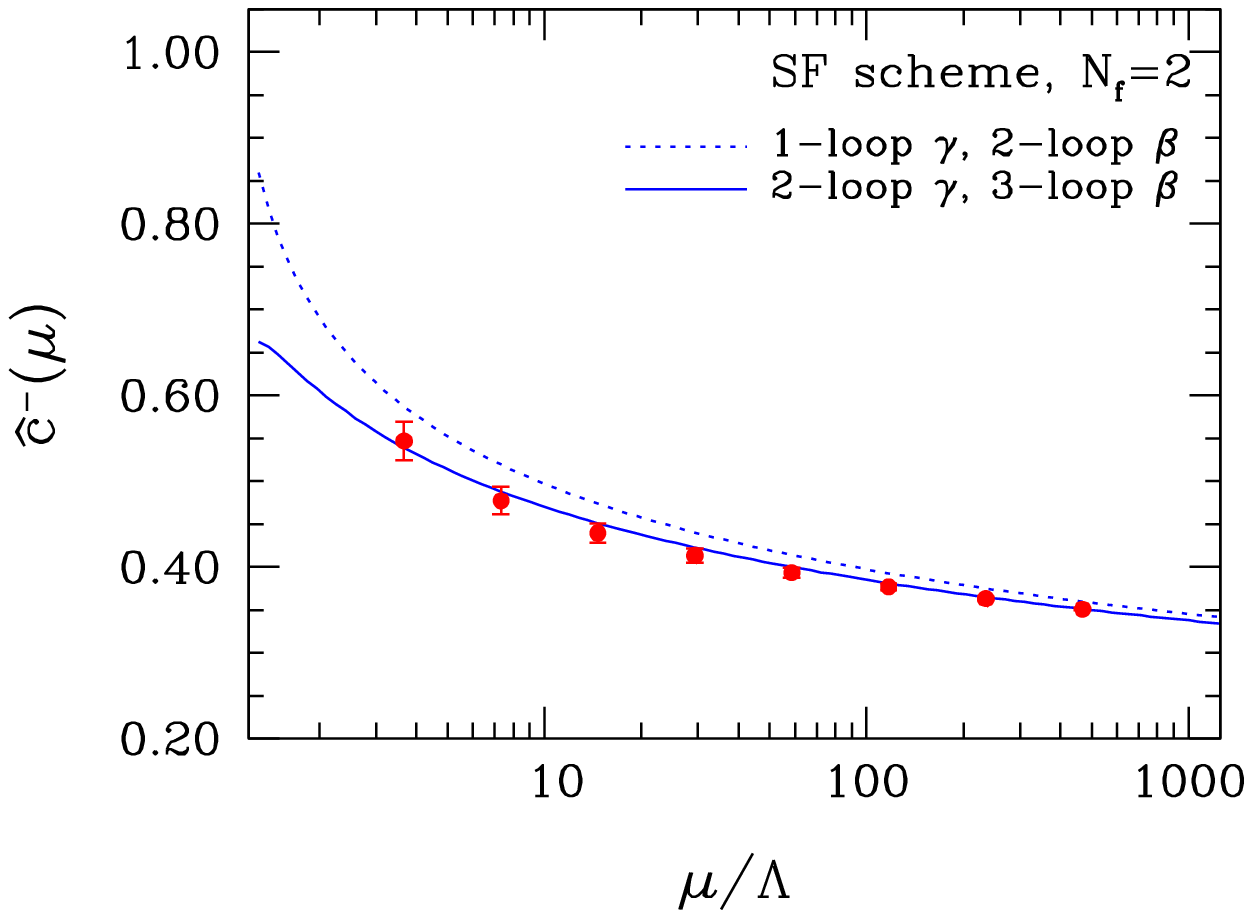, width=6.5 true cm}
  \end{center}
  \vskip -0.7cm
\end{figure}

\begin{figure}[!h]
  \vskip -1.4cm
  \begin{center}
    \vskip 1.5cm
     \hskip -0.2cm \epsfig{figure=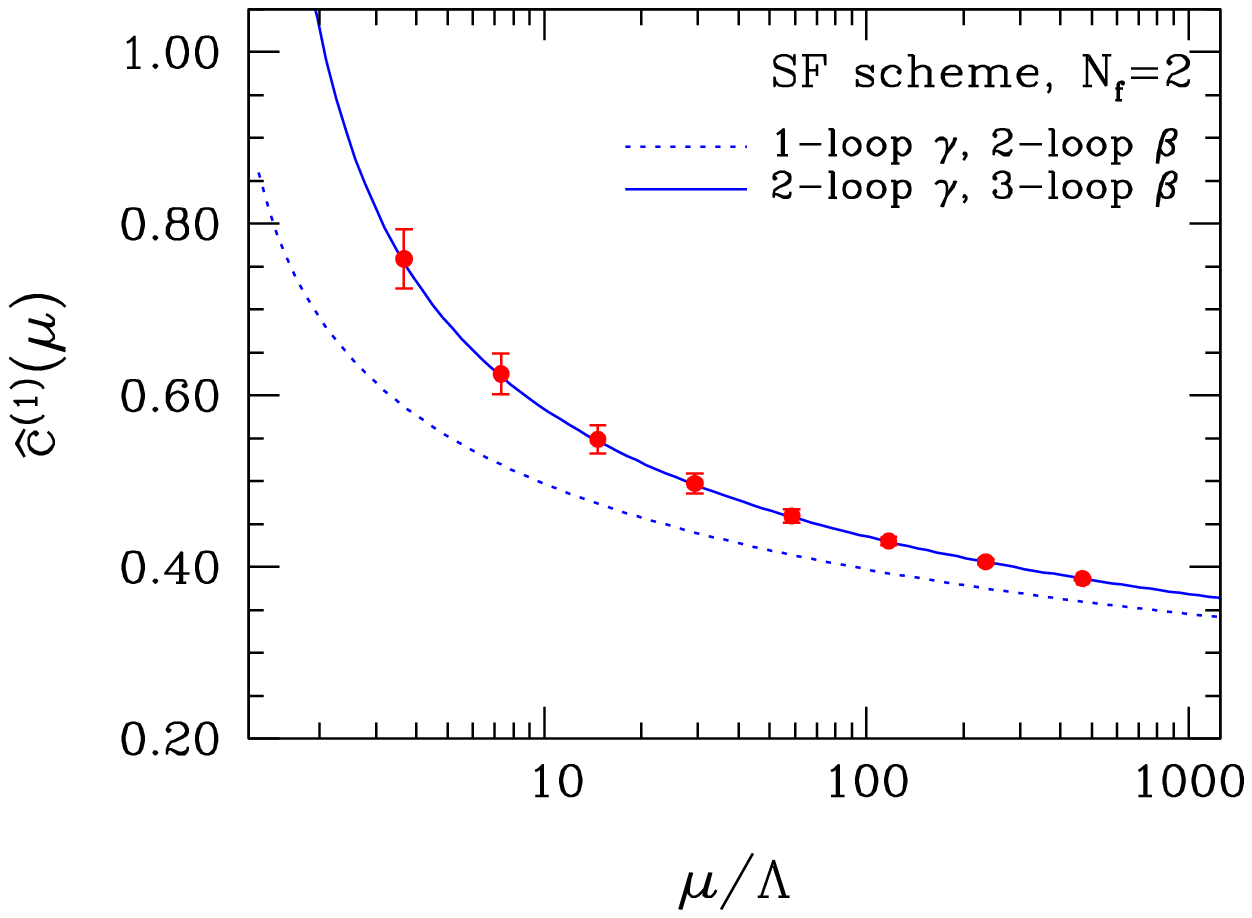, width=6.5 true cm}
     \epsfig{figure=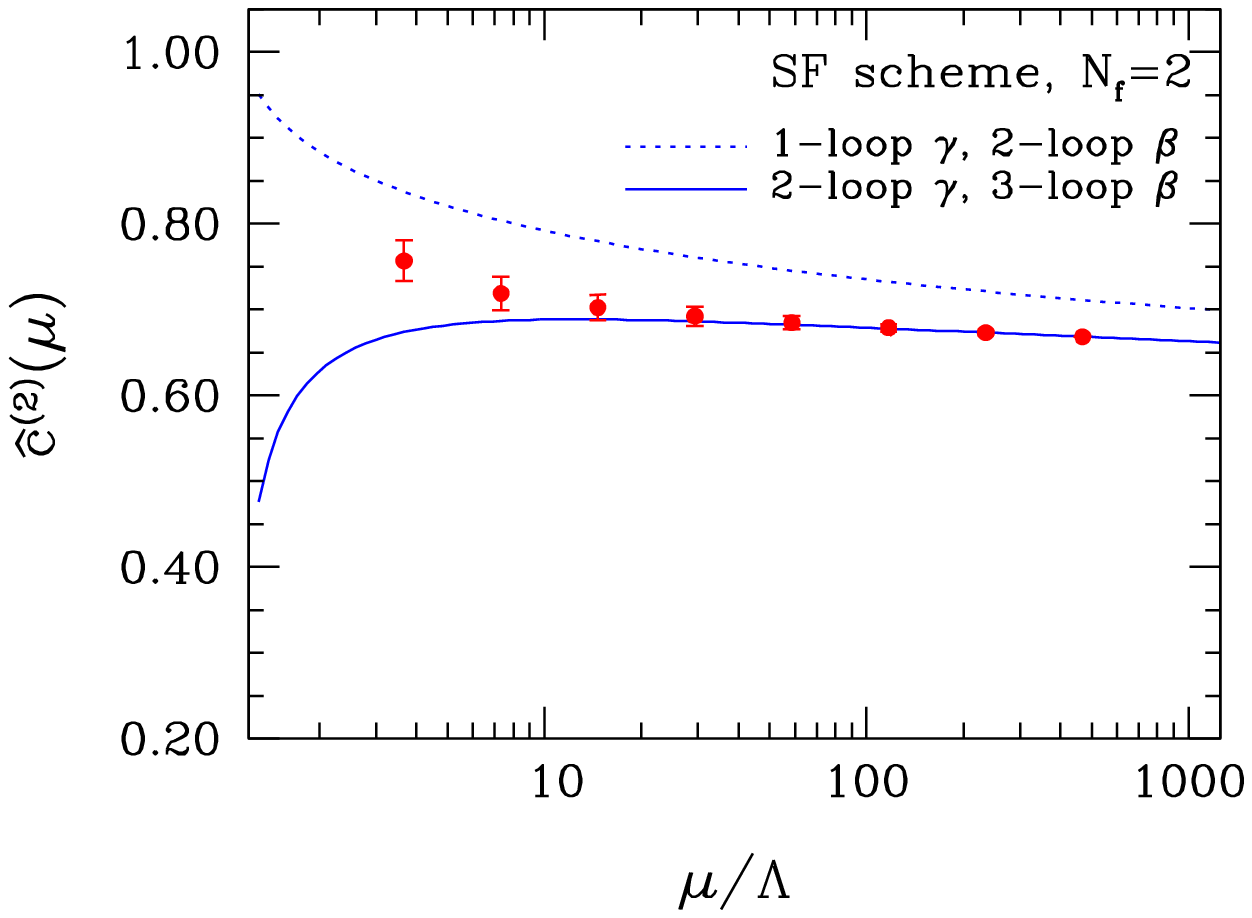, width=6.5 true cm}
  \end{center}
  \vskip -0.7cm
\end{figure}

\end{document}

%% file: macros.tex
\newcommand{\ba}{\begin{array}}
\newcommand{\ea}{\end{array}}

\newcommand{\req}[1]{Eq.~(\ref{#1})}

\newcommand{\dif}{{\rm d}}

\newcommand{\Dslash}{\relax{\kern+.25em / \kern-.70em D}}

\newcommand{\MeV}{{\rm MeV}}

\newcommand{\Real}{\relax{\mathsf{\Gamma\kern-.35em R}}}
\newcommand{\Int}{\relax{\mathsf{Z\kern-.40em Z}}}

\newcommand{\NF}{N_\mathrm{\scriptstyle f}}

\newcommand{\gbar}{\kern1pt\overline{\kern-1pt g\kern-0pt}\kern1pt}
\newcommand{\mbar}{\kern2pt\overline{\kern-1pt m\kern-1pt}\kern1pt}
\newcommand{\obar}[1]{\kern3pt\overline{\kern-2pt #1\kern-0pt}\kern1pt}

\newcommand{\hopc}{\kappa_{\rm cr}}

\newcommand{\lmax}{L_{\rm max}}

\newcommand{\Oa}{\mbox{O}(a)}

\newcommand{\abar}{\kern1pt\overline{\kern-1pt a\kern-0.5pt}\kern1pt}

\newcommand{\cO}{{\cal O}}

\newcommand{\cQ}{{\cal Q}}

\newcommand{\heavy}{\psi_{\rm \phantom{\bar{h}^\dagger}\hspace{-0.28cm}h}}
\newcommand{\heavyb}{{\bar{\psi}}_{\rm h}}
\newcommand{\aheavy}{\psi_{{\rm \phantom{\bar{h}^\dagger}\hspace{-0.28cm}\bar{h}}}}
\newcommand{\aheavyb}{\bar{\psi}_{\bar{\rm h}}}

\newcommand{\bx}{{\mathbf{x}}}
\newcommand{\by}{{\mathbf{y}}}